\documentclass[a4paper,11pt, superscriptaddress, nofootinbib, floatfix,scrartcl, longbibliography]{article}
\pdfoutput=1 

\usepackage{jheppub} 

\usepackage[T1]{fontenc} 
\usepackage{tikz-cd}
\usepackage{mathtools}
\usepackage{verbatim}

\numberwithin{equation}{section}

\usepackage{amssymb}
\usepackage{hyperref}
\usepackage{color}
\usepackage{graphicx}
\usepackage{srcltx}
\usepackage{mathrsfs}
\usepackage{bbold}
\usepackage{amsmath}
\usepackage{slashed}
\usepackage[numbers]{natbib}
\usepackage{filecontents}
\usepackage{atbegshi,picture}
\usepackage{lipsum}
\usepackage{textpos}

\usepackage{tikz}

\usepackage[utf8]{inputenc}

\newcommand{\rn}[1]{%
  \textup{\uppercase\expandafter{\romannumeral#1}}%
}

\AtBeginShipoutNext{\AtBeginShipoutUpperLeft{
  \put(\dimexpr\paperwidth+1cm\relax,-4cm){\makebox[0pt][r]{IPMU-18-0130}}
}}

\title{Discrete gauge anomalies revisited}

\author{Chang-Tse Hsieh}
\affiliation[1]{Kavli Institute for the Physics and Mathematics of the Universe (WPI),
The University of Tokyo Institutes for Advanced Study,
The University of Tokyo, Kashiwa, Chiba 277-8583, Japan}
\affiliation[2]{Institute for Solid State Physics, The University of Tokyo. Kashiwa, Chiba 277-8581, Japan}
\emailAdd{changtse.hsieh@ipmu.jp}

\abstract{
We revisit discrete gauge anomalies in chiral fermion theories in $3+1$ dimensions. We focus on the case that the full symmetry group of fermions is $\mathrm{Spin}(4)\times\mathbb{Z}_n$ or $(\mathrm{Spin}(4)\times\mathbb{Z}_{2m})/\mathbb{Z}_{2}$ with $\mathbb{Z}_2$ being the diagonal $\mathbb{Z}_2$ subgroup. The anomalies are determined by the consistency condition --- based on the Dai-Freed theorem --- of formulating a chiral fermion theory on a generic spacetime manifold with a structure associated with either one of the above symmetry groups and are represented by elements of some finite abelian groups. Accordingly, we give a reformulation of the anomaly cancellation conditions, and 
compare them with the previous result by Ib\'{a}\~{n}ez and Ross. The role of symmetry extensions in discrete symmetry anomalies is clarified in a formal fashion.
We also study gapped states of fermion with an anomalous global $\mathbb{Z}_n$ symmetry, and present a model for constructing these states in the framework of weak coupling.

}

\begin{document} 
\maketitle
\flushbottom

\section{Introduction}
\label{sec:intro}

Cancellation of gauge anomalies is a fundamental constraint on a consistent quantum field theory. For example, in a $(3+1)$-dimensional U(1) chiral gauge theory --- Weyl fermions coupled to a U(1) gauge theory --- the U(1) charges of fermions must satisfy the following relation
\begin{align}
\label{anomaly-free_U(1)}
\Delta q_3 := \sum_Lq_L^3 -\sum_Rq_R^3=0, 
\quad \Delta q_1:= \sum_Lq_L -\sum_Rq_R=0
\end{align}
to ensure the consistency of the theory,  where $q_L$ and $q_R$ are charges of left- and right-handed Weyl fermions, respectively. The first constraint  is required for cancellation of purely gauge anomaly, while the second one is required for cancellation of mixed gauge-gravitational anomaly. Note that the effect of gravity is considered, as the theory should also make sense when coupled to a generic gravitational background. 
These two kinds of anomalies are both perturbative anomalies and can be computed by a conventional Feynman-diagram approach. Alternatively, one can also consider a fermion theory coupled to both a background U(1) gauge field and gravity and require vanishing of the associated 't Hooft anomalies (obstruction of having a well-defined partition function in the above setup), which correspond to the coefficients proportional to $\Delta q_3$ and $\Delta q_1$ of the $(4+1)$-dimensional U(1) gauge and mixed gauge-gravitational Chern-Simons terms, respectively.

While anomalies of continuous symmetries such as U(1) are well understood, the cases of discrete symmetries have been studied much less. Because discrete gauge symmetries can play an important role in constraining the low energy physics of some important theories such as the standard model, the study of discrete gauge anomalies deserves research efforts. Some early results about this issue can be found in \cite{Ibanez-Ross1991, PTWW1991, Banks-Dine1991, Ibanez1992, Csaki:1998aa, Araki2007}. 
One of the arguments for cancellation of discrete gauge anomalies, e.g., in the pioneering work \cite{Ibanez-Ross1991} by Ib\'{a}\~{n}ez and Ross, is based on the cancellation condition of a continuous symmetry in which the low energy discrete symmetry are embedded and on addition constraints regarding the procedure of spontaneous symmetry breaking. 

In this paper, we study discrete gauge anomalies in $(3+1)$d chiral fermion theories from a more modern perspective, based on the concept of symmetry protected topological (SPT) phases. 
In particular, we give a purely low energy description of discrete gauge anomalies ---  as gauge symmetries in many situations are emergent \cite{Witten:2018aa}.
We focus on the simplest case that the discrete internal symmetries are cyclic groups. 
 In this case, the full symmetry group of fermions can be either $\mathrm{Spin}(4)\times\mathbb{Z}_n$ or $\mathrm{Spin}^{\mathrm{Z}_{2m}}(4):=(\mathrm{Spin}(4)\times\mathbb{Z}_{2m})/\mathbb{Z}_{2}$, where $\mathbb{Z}_2$ is the diagonal $\mathbb{Z}_2$ subgroup. 
 
 Since a discrete gauge symmetry is in general associated to massive gauge fields, or a topological gauge theory in the low energy limit, an inconsistency of a chiral gauge theory (coupled to Weyl fermions) usually comes from the presence of non-perturbative or global anomalies.
In this situation, the perturbative Feynman-diagram method might not be useful for studying such anomalies. There should be a topological approach to detect these gauge anomalies; however, we will not take this route in this paper.
Instead, we consider the 't Hooft anomalies of the $\mathrm{Spin}(4)\times\mathbb{Z}_n$ or the $\mathrm{Spin}^{\mathrm{Z}_{2m}}(4)$ symmetry of a chiral fermion theory. That is, we look at the consistency, based on the Dai-Freed theorem \cite{Dai-Freed1994}, of formulating a fermion theory on a generic spacetime manifold endowed with a structure associated with the $\mathrm{Spin}(4)\times\mathbb{Z}_n$ or the $\mathrm{Spin}^{\mathrm{Z}_{2m}}(4)$ group. We explicitly compute these anomalies in the main text, with the main result summarized as follows.

  For a theory of Weyl fermions transforming under the ``untwisted'' symmetry group $\mathrm{Spin}(4)\times\mathbb{Z}_n$, the anomaly-free condition is 
\begin{align}
\label{anomaly-free_untwisted_0}
\left(n^2+3n+2\right) \Delta s_3= 0 \mod 6n, 
\quad 2\Delta s_1=0 \mod n ,
\end{align}
where
$\Delta s_3 := \sum_{L} s_L^3 - \sum_Rs_R^3$ and $\Delta s_1 :=\sum_{L} s_L - \sum_Rs_R$ are defined in terms of the $\mathbb{Z}_n$ charges of fermions that are integers modulo $n$.
On the other hand, for fermions transforming under the ``twisted'' symmetry group $\mathrm{Spin}^{\mathrm{Z}_{2m}}(4)$, the anomaly-free condition is 
\begin{align}
\label{anomaly-free_twisted_0}
(2m^2+m+1)\Delta\tilde{s}_3-(m+3)\Delta\tilde{s}_1=0 \mod 48m,
\quad m\Delta\tilde{s}_3+\Delta \tilde{s}_1= 0 \mod 2m,
\end{align}
where
 $\Delta \tilde{s}_3 := \sum_{L} \tilde{s}_L^3 - \sum_R\tilde{s}_R^3$ and $\Delta \tilde{s}_1 :=\sum_{L} \tilde{s}_L - \sum_R\tilde{s}_R$ are defined in terms of the $\mathbb{Z}_{2m}$ charges of fermions that are odd integers modulo $2m$.  
 
There is an essential difference between the anomaly cancellation condition of a continuous U(1) symmetry and the one of a discrete symmetry: While (\ref{anomaly-free_U(1)}) is independent of the normalization of U(1) charges and of whether the fermions, if all the charges are odd, couple to a U(1) or a $\mathrm{spin}^{c}$ gauge field, (\ref{anomaly-free_untwisted_0}) or (\ref{anomaly-free_twisted_0}) is sensitive to these changes, e.g. a lift from $\mathbb{Z}_n$ to $\mathbb{Z}_{ln}$ or a change from the twisted to the untwisted symmetry --- all are associated to symmetry extensions. For example, let us consider two left-handed Weyl fermions with the same symmetry transformation $\psi_{1,2}\rightarrow i\psi_{1,2}$. Such a theory has a nontrivial anomaly for the $\mathrm{Spin}^{\mathrm{Z}_{4}}(4)$ symmetry, but has no anomaly for an enlarged symmetry $\mathrm{Spin}(4)\times\mathbb{Z}_8$ (so the two fermions can consistently couple to a (topological) $\mathbb{Z}_8$ gauge theory and the total symmetry is extended from $\mathrm{Spin}^{\mathrm{Z}_{4}}(4)$ to $\mathrm{Spin}(4)\times\mathbb{Z}_8$). 
The dependence of discrete anomalies on symmetry extensions is an important issue, which we will discuss in more detail later.

In addition to studying the anomalies of gauge theories, we are also interested in theories with anomalous global symmetries, such as the boundary theories of SPT phases. In some situations, anomalous global symmetries can even be emergent in the low energy phases of a physical system by itself. 
By looking at the 't Hooft anomalies of the global symmetries (incorporating with spacetime symmetry) of a system, we can know some universal properties of the system at low energy, e.g., deformability to a (topologically) trivial/nontrivial gapped phase in a symmetry-preserving fashion.
There have been many discussions and studies along this line in both condensed matter and high energy communities
\cite{Ryu:2012aa, Hsieh:2014aa, Witten2015, You:2015aa, Hsieh:2016aa, Seiberg-Witten2016, Witten2016, Gaiotto:2017aa, Wang:2017aa, Cho:2017aa, Komargodski:2017aa, Tanizaki:2017aa, Shimizu:2018aa, Metlitski-Thorngren17, Kobayashi:2018aa, Yao:2018aa}. 
In this work, we also study gapped states of fermions in a $(3+1)$-dimensional system with an anomalous global $\mathbb{Z}_n$ symmetry. With the help of the anomaly formulas given above, we give an explicit construction for these states.

This paper is organized as follows.

In Section~\ref{Discrete symmetry anomalies in chiral fermion theories in 4d},
we review the modern perspective on ('t Hooft) anomalies of symmetries in fermion theories, which is based on the Dai-Freed theorem, and discuss the related mathematical objects, e.g. the spin bordism/cobordism theory, for the computation of anomalies. We then explicitly calculate the 't Hooft anomalies of $\mathrm{Spin}(4)\times\mathbb{Z}_n$ and $\mathrm{Spin}^{\mathbb{Z}_{2m}}(4)$.

In Section~\ref{Further discussion of anomalies of 4d chiral gauge theory associated with discrete symmetries}, 
we first review the anomaly cancellation conditions argued by Ib\'{a}\~{n}ez and Ross (in the work \cite{Ibanez-Ross1991}), and then discuss the connection between these conditions and the result obtained in Sec.~\ref{Discrete symmetry anomalies in chiral fermion theories in 4d}. In particular, we clarify the role of symmetry extensions on discrete symmetry anomalies.

In Section~\ref{Massive fermions with anomalous global symmetries in 4d},
we studied gapped state of fermions with anomalous global symmetries, from the perspective of anomaly trivialization by symmetry extensions. We present a physical model via weak coupling to realize these states, focusing on the case of untwisted $\mathbb{Z}_n$ symmetries. 


In Appendix~\ref{Details of some derivations}, we fill in the details of the derivation of Eqs.~(\ref{Sn_2}) and (\ref {eta_5d_7d_all_n}).

After finalizing this manuscript, the author became aware of \cite{Garcia-Etxebarria:2018aa}, which partially overlaps our discussion on the anomalies of the $\mathrm{Spin}(4)\times\mathbb{Z}_n$ symmetries as well as the connection to the Ib\'{a}\~{n}ez-Ross conditions.

\section{Discrete symmetry anomalies in chiral fermion theories in 4d}
\label{Discrete symmetry anomalies in chiral fermion theories in 4d}

\subsection{Review of the modern perspective on anomalies of fermions}
 \label{Review of the modern prospective of anomalies of fermions}
 
Consider a set of spin $1/2$ Weyl fermions $\Psi=(\psi_1, \psi_2, \ldots, \psi_N)$ with the same chirality that transform under either the $\mathrm{Spin(4)}\times G$ or the $\mathrm{Spin}^{G}(4):=(\mathrm{Spin(4)}\times G)/\mathbb{Z}_2$ (with $\mathbb{Z}_2$ being the diagonal $\mathbb{Z}_2$ subgroup) group in four spacetime dimensions. Here we work in Euclidean signature and focus on the case that $G$ is a finite group of purely internal symmetries --- not involving spatial-reflection or time-reversal ones. We always assume $G$ has a $\mathbb{Z}_2$ subgroup when $\mathrm{Spin}^{G}(4)$  is involved. 

We formulate the fermion theory on a generic  compact Riemannian four-manifold $(M, g)$ endowed with  a spin structure together with a $G$ structure (denoted as $\mathrm{spin}\times G$ structure) or with a spin$^G$ structure, which we denote as $(M, g, s, f)$, with $M$ being a $\mathrm{spin}\times G$ manifold, or $(M, g, s_G)$, with $M$ being a spin$^G$ manifold, respectively. In the former case, $f$ is a continuous map defined up to homotopy (classifying map) from $M$ to the classifying space $BG$ and $s$ is a spin structure on $TM$ (parametrized by elements of $H^1(M, \mathbb{Z}_2)$).
In the latter case,  
$s_G$ is a spin$^G$ structure on $TM$ that is liftable to a spin$^G$ bundle.
The Weyl fermions are spinors in a section of the twisted spinor bundle $S^{\pm}(M)\otimes V_R$,
where $S^{\pm}(M)$ is the positive/negative spin ``bundle'' over $M$ and  $V_R$ is an associated vector ``bundle'' of the underlying $G$-bundle over $M$ in a representation $R$ of $G$. Here we use the the quotation mark to emphasize that, while $S^{\pm}(M)$ and $V_R$ exist separately on a $\mathrm{spin}\times G$ manifold,  the product $S^{\pm}(M)\otimes V_R$ but not necessarily individual ones exist on a spin$^G$ manifold.
Note that, as $G$ is finite (so $V_R$ is flat), the chiral Dirac operator $\mathcal{D}^{\pm}_R$, which maps sections of $S^{\pm}(M)\otimes V_R$ to sections of $S^{\mp}(M)\otimes V_R$, is locally isomorphic to the form
\begin{align}
 \mathbb{1}_{n\times n}\otimes\left[i\gamma^{\mu} (\partial_{\mu} + \omega_{\mu}) \frac{1\pm\gamma^5}{2} \right],
\end{align}
where $\omega_{\mu}$ is the spin connection and $\gamma^{\mu}$ and $\gamma^5$ are respectively the curved-space gamma matrices and the chirality matrix in four dimensions.

Now we wonder whether the partition function of the system, evaluated as $\det(\mathcal{D}^{\pm}_R)$ on $S^{\pm}(M)\otimes V_R$ by some suitable regularization,
is well-defined or not, in the meaning of respecting gauge and diffeomorphism invariance.
First, since the total Lagrangian density is locally isomorphic to $n=\dim(R)$ copies of the Lagrangian density for a single Weyl fermion without gauge fields, there is no perturbative gravitational (and gauge) anomaly in the theory, while such an anomaly occurs only in $4k+2$ dimensions 
\cite{AG-Witten1984}.
That is, the partition function is invariant under infinitesimal diffeomorphisms. 

However, the theory may have global anomalies, which in general depend on the topology of the bundle $S^{\pm}(M)\otimes V_R$ and are typically more difficult (comparing with the perturbative anomalies) to analyze. 
A traditional definition of global anomalies is given by the non-invariance of the partition function under large diffeomorphisms (or ones combined with gauge transformations if continuous gauge fields are present). These anomalies are represented by $\mathrm{U}(1)$ phases that can be evaluated by the (exponentiated) $\eta$ invariants of the five-dimensional Dirac operator on all possible twisted spinor bundles (for a given representation $R$ of $G$) over the mapping tori obtained by gluing together the ends of $M\times [0, 1]$ via large diffeomorphisms that preserve all the relevant structures on $M$
\cite{Witten1982, Witten1985}.

Yet this is still not the whole story for the problem of anomalies. If there exists any five-dimensional manifold with boundary $M$ such that all the metric and the  $\mathrm{spin}\times G$ or spin$^G$ structure on $M$ can extend over it,
the theory on $M$, as a boundary theory of some theory defined on the five-manifold, should not depend on the way it extends in one dimension higher. To be more specific, let $X$ be a five-manifold with boundary $\partial X=M$. Then the Dai-Freed theorem 
\cite{Dai-Freed1994}
gives a physically sensible definition of the partition function of the whole system  
\cite{Witten2015, Witten2016, Yonekura2016}
\begin{align}
\label{Dai-Freed_theorem}
 Z_{\Psi}= |\det \mathcal{D}^{\pm}_R(M)| \exp(-2\pi i\eta_R(X)).
\end{align}
Here 
$\eta_R(X)$ is the Atiyah-Patodi-Singer (APS)  $\eta$ invariant of the Dirac operator $\mathcal{D}_R(X)$ on a twisted spinor bundle over $X$ that equals $S^{\pm}(M)\otimes V_R$ --- on which the chiral Dirac operator $\mathcal{D}^{\pm}_R(M)$ acts ---  when restricted to $M$;
it is defined as an analytic measure of the spectral asymmetry of $\mathcal{D}_R(X)$:
\begin{align}
\label{Dai-Freed_thm}
 \eta_R(X)&=\frac{1}{2}\lim_{s \rightarrow 0}\left(
 \sum_{\lambda \neq 0}\text{sign}(\lambda)\cdot |\lambda|^{-s}
 + \dim\text{Ker}\left(\mathcal{D}_R(X)\right)
  \right),  
\end{align}
where $\lambda$ are nonzero eigenvalues of $\mathcal{D}_R(X)$ and a regularization of the infinite sum at $s=0$ is taken.
Then, we would like to ask if the formula (\ref{Dai-Freed_theorem}) depends on the twisted spinor bundle (over a five-manifold) on which $\eta_R$ is evaluated. If so, the theory of massless fermions $\Psi$ on $(M, g, s, f)$ or $(M, g, s_G)$, with the partition function defined via the formula (\ref{Dai-Freed_theorem}), is anomalous, in the sense of being a purely four-dimensional theory;
that is, $Z_{\Psi}$ includes the contribution from the (bulk) partition function in five dimensions. 
This is a refined definition of the global anomalies given in 
\cite{Witten2015, Witten2016}.

\begin{figure}[tbp]
\centering
\scalebox{0.7}{\includegraphics{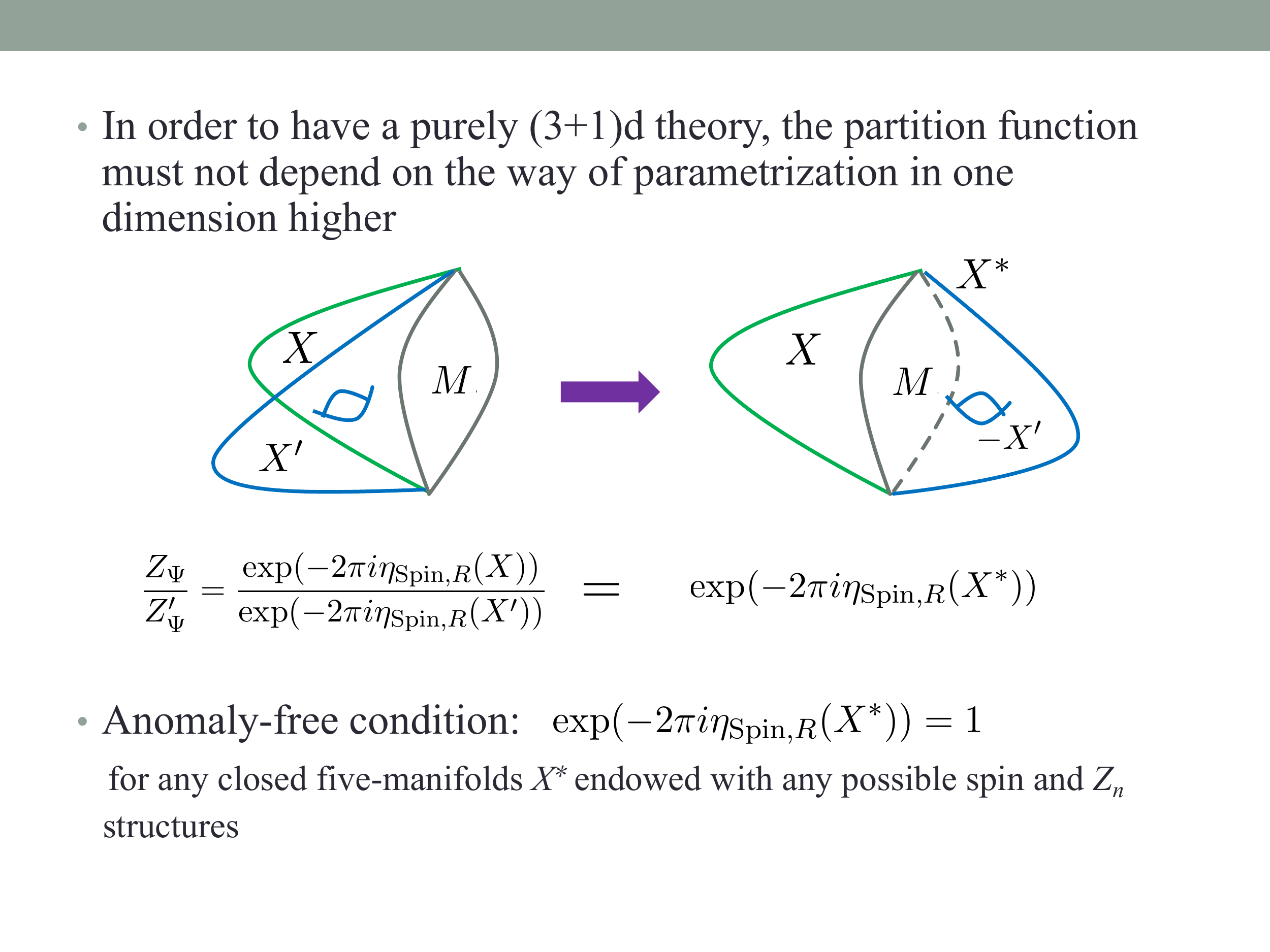}}
\caption{
\label{glunglaw_eta_inv}
Two manifolds $X$ and $X'$ with the same boundary $M$ are glued together along $M$ to make a closed manifold $X^*$.
}

\end{figure}

The condition for whether a theory is free from such kind of anomalies can be determined in the following way. Suppose there exist two five-manifolds $X$ and $X'$ with the same boundary $M$ such that the metric and all the structures on $M$ extend over each of them. The two twisted spinor bundles over $X$ and $X'$ both coincide with $S^+(M)\otimes V_R$ when restricted to $M$.
By reversing the orientation of $X'$ and by taking an appropriate $\mathrm{spin}\times G$ or spin$^G$  structure associated with this reversal,  one can then glue $X$ and $X'$ (and all their structures) together along $M$ to make a closed manifold $X^* = X\cup (-X')$,
as shown in FIG.\ \ref{glunglaw_eta_inv}.
Since the $\eta$ invariant respects a gluing law as the usual gluing relation for any local effective action on manifolds (and bundles)
\cite{Witten2016, Dai-Freed1994}, 
one has
\begin{align}
 \frac{Z_{\Psi}}{Z'_{\Psi}}
&= \frac{\exp(-2\pi i\eta_R(X))}{\exp(-2\pi i\eta_R(X'))}
\nonumber\\
& =\exp(-2\pi i\eta_R(X))\exp(-2\pi i\eta_R(-X'))
\nonumber\\
 &=\exp(-2\pi i\eta_R(X^*)).
\end{align}
Now it is obvious that $Z_{\Psi}$ given by the formula (\ref{Dai-Freed_theorem}) does not depend on the choice of $X$ as well as the $\mathrm{spin}\times G$ or spin$^G$ structure on it if and only if $\exp(-2\pi i\eta_R(X^*))$ equals 1 on any closed five-manifolds $X^*$ together with the associated structures.

As $G$ is finite,  $\exp(-2\pi i\eta_R(X^*))$ or $\eta_R(X^*) \mod \mathbb{Z}$ is a bordism invariant.
That is, if $X^*$  bounds a six-dimensional spin manifold $Z$ such that all the structures on $X^*$ extend over $Z$, the APS index theorem 
\cite{Atiyah-Patodi-Singer1975, Gilkey}
tells us that $\eta_R(X^*)$ equals the index of the Dirac operator on the twisted spinor bundle over $Z$ (with the APS boundary condition) and thus $\eta_R(X^*)$ is an integer. Note that there is no contribution from the local invariant in the bulk of $Z$ to this index, because the Dirac genus of $Z$, $\hat{A}(Z)$, vanishes in six dimensions. (This also means there is no perturbative gravitational anomalies in the four-dimensional fermion theory, as mentioned before.)

The Dai-Freed theorem gives a natural way to ``classify'' the anomaly of the four-dimensional massless fermions $\Psi$ in an arbitrary ordinary $G$-representation or spin$^G$ representations $R$, through the $\eta$ invariant map 
\begin{align}
\eta_R: \Omega^{\mathrm{Spin}}_5(BG) \rightarrow \mathbb{R}/\mathbb{Z}\quad
\text{by}\quad
[(X^*, g, s, f)] \mapsto \eta_R(X^*) \mod \mathbb{Z},
\end{align}
or
\begin{align}
\eta_R: \Omega^{\mathrm{Spin}^G}_5\rightarrow \mathbb{R}/\mathbb{Z}\quad
\text{by}\quad
[(X^*, g, s_G)] \mapsto \eta_R(X^*) \mod \mathbb{Z},
\end{align}
where $\Omega^{\mathrm{Spin}}_5(BG)$ and $\Omega^{\mathrm{Spin}^G}_5$ are the bordism groups
of closed five-manifolds with $\mathrm{spin}\times G$ and spin$^G$ structures, respectively.
We denote elements of 
$\Omega^{\mathrm{Spin}}_5(BG)$ / $\Omega^{\mathrm{Spin}^G}_5$ by the bordism classes of topological spaces $[(X^*, g, s, f)]$ / $[(X^*, g, s_G)]$. 
Furthermore, $\eta_R\mod\mathbb{Z}$ or its exponential $\exp(-2\pi i\eta_R)$ is also regarded as an element of the fermionic SPT phases with $G$ in five dimensions.  The U(1)-valued topological (bordism) invariant $\exp(-2\pi i\eta_R(X^*))$ is the partition function of an invertible topological quantum field theory (TQFT), which describes a fermionic SPT phase at  low energy, on a closed five-dimensional $\mathrm{spin}\times G$ or spin$^G$ manifold $X^*$.

It has been proposed that fermionic SPT phases with a generic (untwisted) symmetry group $G$ in $d$ dimensions can be classified by elements of the group 
$\mathrm{Hom}(\Omega^{\mathrm{Spin}}_{d,\mathrm{tors}}(BG), \mathrm{U}(1))$,
where $\Omega^{\mathrm{Spin}}_{d,\mathrm{tors}}(BG)$ is the torsion subgroup of $\Omega^{\mathrm{Spin}}_{d}(BG)$, the d-dimentional spin bordism group \cite{Kapustin2014c}.
(See also some recent discussions \cite{Xiong:2016aa, Yonekura:2018aa}.)
For $d=5$ and $G$ being a finite group, $\Omega^{\mathrm{Spin}}_{5,\mathrm{tors}}(BG) = \Omega^{\mathrm{Spin}}_{5}(BG)$,
and the exponential $\eta$ invariant maps $\exp(-2\pi i\eta_R)$ for all representations of $G$ generate a subgroup of the spin cobordism group
$\Omega^{5}_{\mathrm{Spin}}(BG) :=\mathrm{Hom}(\Omega^{\mathrm{Spin}}_{5}(BG), \mathrm{U}(1))$, 
\footnote{
A more formal use of the the notation $\Omega^{*}_{\mathrm{Spin}}(X)$ is in a generalized cohomology theory with $\Omega_{*}^{\mathrm{Spin}}(X)$ as a generalized homology group and $\Omega^{*}_{\mathrm{Spin}}(X)$ as a generalized cohomology for the space $X$. Here we just defined $\Omega^{*}_{\mathrm{Spin}}(X)$ as the homomorphism of the bordism group $\Omega^{*}_{\mathrm{Spin}}(X)$ for convenience.
} 
which we denote as
\begin{align}
\Gamma^{5}_{\mathrm{Spin}}(BG) \subseteq 
\Omega^{5}_{\mathrm{Spin}}(BG).
\end{align}
As discussed above, elements of $\Gamma^{5}_{\mathrm{Spin}}(BG)$ correspond to SPT phases of free fermions in five dimensions, and thus, through the bulk-boundary correspondence, classify the anomalies of theories of massless fermions with symmetry $G$ in four dimensions.
It is clear an anomaly-free representation of $G$ corresponds to the identity element of $\Gamma^{5}_{\mathrm{Spin}}(BG)$.

In general, there might exist manifolds with spin and $G$ structures that, as nontrivial elements of $\Omega^{\mathrm{Spin}}_{5}(BG)$, can not be detected by 
$\exp(-2\pi i\eta_R)$ for any representations of $G$; that is, $\exp(-2\pi i\eta_R)$  equals 1 when evaluated on these manifolds. In this case, 
$\Gamma^{5}_{\mathrm{Spin}}(BG)$ is a proper subgroup of $\mathrm{Hom}(\Omega^{\mathrm{Spin}}_{5}(BG), \mathrm{U}(1))$; elements of the latter but not of the former correspond to SPT phases that can not be described by free fermions. 

The above statement on the relation between fermionic SPT phases with untwisted symmetries and cobordism groups should also apply to the case of twisted symmetries, so we would not repeat the discussion on the latter. 

In this paper, we focus on the case that $G$ is a cyclic group (which is abelian). We denote $G=\mathbb{Z}_n$ for the untwisted case and $G=\mathbb{Z}_{2m}$ for the twisted case, respectively.
In the twisted case, an alternative description of the spin$^{\mathbb{Z}_{2m}}$ bordism group $\Omega^{\mathrm{Spin}^{\mathbb{Z}_{2m}}}_5$ is given by the ``twisted'' spin bordism group $\Omega^{\mathrm{Spin}}_5(B\mathbb{Z}_m, \xi)$ via a real vector bundle $\xi$ over $B\mathbb{Z}_m$ with a vanishing first Stiefel-Whitney class (which is true as we are focusing on symmetries preserving the orientation of spacetime) and a map $f: M \rightarrow B\mathbb{Z}_m$ such that a spin$^{\mathbb{Z}_{2m}}$ structure on $TM$ corresponds to a spin structure on $TM\oplus\xi$
\cite{Botvinnik:1997aa, Barrera-Yanez:2006aa}.
Moreover, if the second Stiefel-Whitney class $w_2\in H^2(B\mathbb{Z}_m, \mathbb{Z}_2)$ of $\xi$ also vanishes (which means $\xi$ is spin), $\Omega^{\mathrm{Spin}}_5(B\mathbb{Z}_m, \xi)$ can be further identified by the untwisted one $\Omega^{\mathrm{Spin}}_5(B\mathbb{Z}_m)$. 
\footnote{
Actually, for a generic twisted spin bordism group $\Omega^{\mathrm{Spin}}_5(BG, \xi)$, any spin structure on $\xi$ gives an isomorphism between
$\Omega^{\mathrm{Spin}}_5(BG)$ and $\Omega^{\mathrm{Spin}}_5(BG, \xi)$
\cite{Botvinnik:1997aa}. 
}
Since $H^2(B\mathbb{Z}_m, \mathbb{Z}_2)=\mathbb{Z}_2$ for even $m$ and $H^2(B\mathbb{Z}_m, \mathbb{Z}_2)=0$ for odd $m$, we have the fact that $\Omega^{\mathrm{Spin}^{\mathbb{Z}_{2m}}}_5 = \Omega^{\mathrm{Spin}}_5(B\mathbb{Z}_m)$ for any odd $m$.

In the following section, we will compute the groups $\Gamma^{5}_{\mathrm{Spin}}(B\mathbb{Z}_n)$ and $\Gamma_{\mathrm{Spin}^{\mathbb{Z}_{2m}}}^5$ (and all the corresponding bordism/corbordism groups) to study the discrete symmetry anomalies of fermions. 

%
%

\subsection{'t Hooft anomalies of $\mathrm{Spin}(4)\times\mathbb{Z}_n$}
\label{'t Hooft anomalies of Spin(4)xZn}

In this section we consider Weyl fermions transforming under the $\mathrm{Spin}(4)\times\mathbb{Z}_n$ group. We compute the group $\Gamma^{5}_{\mathrm{Spin}}(B\mathbb{Z}_n)$ and determine the 't Hooft anomaly $\alpha_R$ (defined later) of $\Psi$ in an arbitrary representation $R$ of $\mathbb{Z}_n$. 
We denote the $\eta$ invariant $\eta_R(X)$ on a closed five-dimensional spin$\times\mathbb{Z}_n$ manifold $X$ as $\eta(X, R)$, where we put ``$R$'' into the parentheses to avoid messy indices when doing ring operations on representations of $\mathbb{Z}_n$.

Let $\mathbb{Z}_n =\{\lambda\in \mathbb{C} : \lambda^n =1 \}$ be the cyclic group of order $n$. Let $\rho_s (\lambda) =\lambda^s$ be a one-dimensional representation of $\mathbb{Z}_n$, where $s$ is an integer defined modulo $n$. Any representation $R$ of $\mathbb{Z}_n$ is an element of the unitary group representation ring of $\mathbb{Z}_n$:
\begin{align}
RU(\mathbb{Z}_n)=\oplus_{s=0}^{n-1} \rho_s\cdot\mathbb{Z}.
\end{align}
We also identify $\mathbb{Z}_n=\mathbb{Z}/n\mathbb{Z}$ by sending $s$ to $e^{2\pi i s/n}$. This gives $\mathbb{Z}_n$ the structure of a ring.

To compute $\Gamma^{5}_{\mathrm{Spin}}(B\mathbb{Z}_n)$, we need to consider all bordism classes -- as  $\exp(-2\pi i\eta(X, R))$ or $\eta(X, R) \mod \mathbb{Z}$  are bordism invariants -- of five-dimensional spin manifolds with $\mathbb{Z}_n$ structures that can be detected by free fermions, which are described by the Dirac theory.
These classes form a group which we denoted as $\Gamma^{\mathrm{Spin}}_5(B\mathbb{Z}_n)$
and are defined through the following equivalence relation:
If $X_1$ and $X_2$ are two five-dimensional spin manifolds endowed with $\mathbb{Z}_n$ structures, $X_1\sim X_2$ if $\eta(X_1 - X_2, R)=0 \mod \mathbb{Z}$ for all representations $R\in RU(\mathbb{Z}_n)$, and we denote $[X]_{\eta}\in\Gamma^{\mathrm{Spin}}_{5}(B\mathbb{Z}_n)$ be the equivalence class of $X$ associated with this equivalence relation.
Clearly, the group $\Gamma^{5}_{\mathrm{Spin}}(B\mathbb{Z}_n)$ defined previously is the Pontryagin dual of $\Gamma^{\mathrm{Spin}}_5(B\mathbb{Z}_n)$, that is,
$\Gamma^{5}_{\mathrm{Spin}}(B\mathbb{Z}_n) = \mathrm{Hom}(\Gamma^{\mathrm{Spin}}_5(B\mathbb{Z}_n), \mathrm{U}(1))$.
In general, it is not easy to determine the group $\Gamma^{\mathrm{Spin}}_d(BG)$ (and $\Omega^{\mathrm{Spin}}_{d}(BG)$) for a generic group $G$ in arbitrary dimensions. Nevertheless, the result in 
\cite{BarreraYanes-Gilkey1999} 
gives a way to evaluate $\Gamma^{\mathrm{Spin}}_d(B\mathbb{Z}_n)$ in terms of the representation theory of $\mathbb{Z}_n$, and we will follow their construction to compute $\Gamma^{\mathrm{Spin}}_5(B\mathbb{Z}_n)$ in this section. Moreover, as their original result focused only on the case $n=2^v$, we also generalize it to any integer $n$.

Following 
\cite{BarreraYanes-Gilkey1999},
we consider the lens space bundles (over $S^2$), a class of five-dimensional spin manifolds endowed with nontrivial $\mathbb{Z}_n$ structures, to study the $\eta$ invariants. They are quotients of the unit sphere bundle of the Whitney sum of the tensor square of the complex Hopf line bundle $H$ and a trivial complex line bundle $\mathbb{1}$ over $S^2$
\begin{align}
\label{lens_space_bundele}
X(n; a_1, a_2) := S(H\otimes H\oplus \mathbb{1})/ \tau(a_1, a_2),
\end{align}
where $\tau(a_1, a_2) := \rho_{a_1}\oplus\rho_{a_2}$ is a representation of $\mathbb{Z}_n$ in $\mathrm{U}(2)$ and its action (by multiplication by $\lambda^{a_i}$ on the $i$-th summand) on the associated unit sphere bundle is fixed-point free, that is, $a_1$ and $a_2$ are both coprime to $n$. By construction, the lens space bundles inherit natural spin structures and $\mathbb{Z}_n$ structures by the identification $\pi_1(X(n; a_1, a_2))=\mathbb{Z}_n$. $X(n; a_1, a_2)$ has a unique spin structure if $n$ is odd, while it has two inequivalent spin structures if $n$ is even; we fix the spin structure for even $n$ by taking the positive sign of the square root of the determinant line bundle $\det(\rho_{a_1}\oplus\rho_{a_2})$. 

The $\eta$ invariant on the lens space bundles can be computed by the following combinatorial formula
\cite{Botvinnik-Gilkey-Stolz1997, BarreraYanes-Gilkey1999}
\begin{align}
\label{combinatorial_eta}
\eta(X(n; a_1, a_2), R) =  \frac{1}{n} \sum_{\lambda\in\mathbb{Z}_n, \lambda\neq 1} \mathrm{Tr}(R(\lambda))
\frac{\lambda^{\frac{1}{2}(a_1+a_2)}(1+\lambda^{a_1})}{(1-\lambda^{a_1})^2(1-\lambda^{a_2})},
\end{align}
where $R\in\ RU(\mathbb{Z}_n)$. Using the $\eta$ invariant, one can construct isomorphisms from some additive abelian groups formed by spanning sets (over $\mathbb{Z}$) of lens space bundles, which are subgroups of $\Gamma^{\mathrm{Spin}}_5(B\mathbb{Z}_n)$, to the representation theory of $\mathbb{Z}_n$, from which these Abelian subgroups can be further represented in terms of cyclic groups.
Here we present the main result about these isomorphisms and leave the proof in the appendix.

We first consider the case that $n$ is a prime power, and then generalize the result to any integer $n$.
In the following discussion, when we write $n=p^v$, we implicitly assume $p$ is a prime number. 
Let
\begin{align}
\label{Sn_1}
S_n :=
\left\{
\begin{array}{ll}
\mathrm{span}_{\mathbb{Z}}\{[X(n; 1, 1)]_{\eta}\},
\quad&\text{if}\ n=2, 3,
 \\
\mathrm{span}_{\mathbb{Z}}\{[X(n; 1, 1)]_{\eta}, [X(n; 1, 3)]_{\eta}\},
\quad &\text{if}\ n=2^v> 2,
\\ 
\mathrm{span}_{\mathbb{Z}}\{[X(n; 1, 1)]_{\eta}, [X(n; 1, 5)]_{\eta}\},
\quad &\text{if}\ n=3^v>3,
\\ 
\mathrm{span}_{\mathbb{Z}}\{[X(n; 1, 1)]_{\eta}, [X(n; 1, 3)]_{\eta}\},
\quad &\text{if}\ n=p^v,\  p>3.
 \end{array}
\right.
\end{align}
By construction, $S(n)\subseteq\Gamma^{\mathrm{Spin}}_5(B\mathbb{Z}_n)$ for each $n=p^v$. 
Then, as shown in Appendix \ref{isomorphisms}, $S_n$ can be expressed by the representation theory of $\mathbb{Z}_n$:
\begin{align}
\label{Sn_2}
S_n \cong I_n/\{I_n\cap RU_0(\mathbb{Z}_n)^4\},
\quad \forall n=p^v,
\end{align}
where
\begin{align}
\label{In}
I_n =\oplus_{j\ge0}(\rho_1-\rho_{-1})(\rho_0-\rho_1)^{2j}\rho_{-1}^j\cdot\mathbb{Z}
\end{align}
and $RU_0(\mathbb{Z}_n)$ is the argumentation ideal of representations of $\mathbb{Z}_n$ with virtual dimension 0, that is,
\begin{align}
RU_0(\mathbb{Z}_n) = (\rho_1-\rho_0)\cdot RU(\mathbb{Z}_n).
\end{align}

The representation theory $I_n/\{I_n\cap RU_0(\mathbb{Z}_n)^4\}$ of $\mathbb{Z}_n$ for each $n=p^v$ can be identified in terms of (direct sums of) cyclic groups.
The case of $n=2$ is trivial as $I_2=0$. For $n=p^v >2$, we express 
$I_n=\oplus_{j\ge0}(\rho_1-\rho_{-1})(\rho_0-\rho_1)^{2j}\rho_{-1}^j\cdot\mathbb{Z}=\oplus_{s=1}^{n-1}(\rho_s-\rho_{-s})\cdot\mathbb{Z}$ modulo the condition $(\rho_s-\rho_0)^4=1$ (as well as $\rho_s^n=1$),  writing $x=\rho_1-\rho_0 \in RU_0(\mathbb{Z}_n)$, as
\begin{align}
&I_n/\{I_n\cap RU_0(\mathbb{Z}_n)^4\}
\nonumber\\
&=
\{(\rho_1-\rho_{-1})\cdot\mathbb{Z}\oplus(\rho_2-\rho_{-2})\cdot\mathbb{Z}\}
/\{I_n\cap RU_0(\mathbb{Z}_n)^4\}
\nonumber\\
&=
\{\{(1+x)-(1+x)^{n-1}\}\cdot\mathbb{Z}\oplus\{(1+x)^2-(1+x)^{n-2}\}\cdot\mathbb{Z}\}
/\{\{(1+x)^n-1\}\cdot\mathbb{Z}[x]+x^4\cdot\mathbb{Z}[x]\}
\nonumber\\
&=
\left\{
\left\{(n-2)x
+\left(
 \begin{smallmatrix}
n-1\\ 2
\end{smallmatrix}
\right) 
x^2+
\left(
 \begin{smallmatrix}
n-1\\ 3
\end{smallmatrix}
\right) 
x^3\right\}\cdot\mathbb{Z}
\oplus
\left\{(n-4)x
+\left[\left(
 \begin{smallmatrix}
n-2\\ 2
\end{smallmatrix}
\right) -1\right]
x^2+
\left(
 \begin{smallmatrix}
n-2\\ 3
\end{smallmatrix}
\right) 
x^3\right\}\cdot\mathbb{Z}
\right\}
\nonumber\\
&\quad\
/\left\{
\left\{nx
+\left(
 \begin{smallmatrix}
n\\ 2
\end{smallmatrix}
\right) 
x^2+
\left(
 \begin{smallmatrix}
n\\ 3
\end{smallmatrix}
\right) 
x^3\right\}\cdot\mathbb{Z}[x]
+
x^4\cdot\mathbb{Z}[x]
\right\}
\end{align}
By computation, we found
\begin{align}
\label{iso_In_cyclic}
I_n/\{I_n\cap RU_0(\mathbb{Z}_n)^4\}
&\cong
\left\{
\begin{array}{ll}
0,
\quad &\text{if}\ n=2,
 \\
\mathbb{Z}_{n} \oplus \mathbb{Z}_{n/4},
\quad &\text{if}\ n=2^v>2,
\\
\mathbb{Z}_{3n} \oplus \mathbb{Z}_{n/3},
\quad &\text{if}\ n=3^v,
\\ 
\mathbb{Z}_{n} \oplus \mathbb{Z}_{n},
\quad &\text{if}\ n=p^v,\ p>3.
 \end{array}
\right.
\end{align}


One can further identify the abelian groups $S_n$ with the spin bordism groups  $\Omega^{\mathrm{Spin}}_{5}(B\mathbb{Z}_n)$ for these values of $n$ by using some spectral sequences that give upper bounds for the orders of $\Omega^{\mathrm{Spin}}_{5}(B\mathbb{Z}_n)$. 
For $n=2^v$, $|\Omega^{\mathrm{Spin}}_{5}(B\mathbb{Z}_n)|$ are estimated by the Adams spectral sequence
\cite{Gilkey-Botvinnik1996}:
\begin{align}
|\Omega^{\mathrm{Spin}}_{5}(B\mathbb{Z}_{n})|\le n^2/4,
\quad n=2^v,\ v\ge 1.
\end{align}
For $n=p^v$ of any odd prime $p$, $|\Omega^{\mathrm{Spin}}_{5}(B\mathbb{Z}_n)|$ are estimated by the Atiyah-Hirzebruch spectral sequence
\cite{Gilkey}:
\begin{align}
&|\Omega^{\mathrm{Spin}}_{5}(B\mathbb{Z}_{n})|
\le \prod_{a+b=5}|\tilde{H}_a(B\mathbb{Z}_n, \Omega^{\mathrm{Spin}}_b(pt))|
\nonumber\\
&= |\tilde{H}_0(B\mathbb{Z}_n, 0)| \cdot |\tilde{H}_1(B\mathbb{Z}_n, \mathbb{Z})| \cdot |\tilde{H}_2(B\mathbb{Z}_n, 0)|
\cdot |\tilde{H}_3(B\mathbb{Z}_n, \mathbb{Z}_2)| \cdot |\tilde{H}_4(B\mathbb{Z}_n, \mathbb{Z}_2)| \cdot |\tilde{H}_5(B\mathbb{Z}_n, \mathbb{Z})|
\nonumber\\
&=1 \cdot |\mathbb{Z}_n| \cdot 1 \cdot 1 \cdot 1 \cdot |\mathbb{Z}_n|
\nonumber\\
&=n^2,
\quad n=p^v,\ v\ge 1,
\end{align}
where $\tilde{H}_k(B\mathbb{Z}_n, M)$ are the reduced homology groups
 of $B\mathbb{Z}_n$ with coefficients in an abelian group $M$.
Observing the order of $S_n$ for each $n=p^v$ in (\ref{iso_In_cyclic}), we then conclude, 
\begin{align}
\label{S(n)_3}
S_n = \Gamma^{\mathrm{Spin}}_5(B\mathbb{Z}_n) = \Omega^{\mathrm{Spin}}_{5}(B\mathbb{Z}_{n}),
\quad  \forall n=p^v,
\end{align}
according to the definitions of these groups.

  Therefore, the (classes of) lens space bundles $[X(n; 1, a)]_{\eta}$ for $a=1, 3, 5$, depending on the value of $n=p^v$, are generators (which are not unique) of $\Gamma^{\mathrm{Spin}}_5(B\mathbb{Z}_n)$ and also $\Omega^{\mathrm{Spin}}_{5}(B\mathbb{Z}_{n})$. In this case, we can identify the equivalence classes $[\ \cdot\ ]_{\eta}$ with the bordism classes $[\ \cdot\ ]$.

The result obtained so far can be generalized to any positive integers $n$. This is essentially based on the the following property of bordism groups 
\cite{Gilkey}
\begin{align}
\label{Omega_coprime_mn}
 \Omega^{\mathrm{Spin}}_{5}(B\mathbb{Z}_{mn}) \cong 
 \Omega^{\mathrm{Spin}}_{5}(B\mathbb{Z}_{m})\oplus  \Omega^{\mathrm{Spin}}_{5}(B\mathbb{Z}_{n}),
\quad \text{if}\ \gcd(m, n)=1.
\end{align}
Clearly,  the abelain groups $S(n)$ and $\Gamma^{\mathrm{Spin}}_5(B\mathbb{Z}_n)$ also satisfy the above property.
Similarly, the relation (\ref{Sn_2}) also extends to any positive integer $n$, that is 
\begin{align}
\label{Gamma_iso_rep_theory}
\Omega^{\mathrm{Spin}}_{5}(B\mathbb{Z}_{n})=\Gamma^{\mathrm{Spin}}_5(B\mathbb{Z}_n) \cong I_n/\{I_n\cap RU_0(\mathbb{Z}_n)^4\}
\cong\mathbb{Z}_{a_n} \oplus \mathbb{Z}_{b_n},
\quad\forall n\in\mathbb{N}.
\end{align}
Here the integers $a_n$ and $b_n$ are defined, when expressing $n=2^{p}\cdot 3^{q}\cdot k^{r}$ with $k\ge 5$ being an odd number not divisible by 3 and $p, q, r$ being nonnegative integers, as
\begin{align}
\label{def_an_bn}
a_n &:=
\left\{
\begin{array}{ll}
k^r,
\quad &\text{if}\ p=0, 1\ \&\ q=0,
 \\
3^{q+1}\cdot k^r,
\quad &\text{if}\ p=0, 1\ \&\ q\ge 1,
\\
2^{p}\cdot k^r,
\quad &\text{if}\  p\ge 2\ \&\ q=0,
\\ 
2^p\cdot 3^{q+1}\cdot k^r,
\quad &\text{if}\ p\ge 2\ \&\ q\ge 1,
 \end{array}
\right.
\nonumber\\
b_n &:=
\left\{
\begin{array}{ll}
k^r,
\quad &\text{if}\ p=0, 1\ \&\ q=0,
 \\
3^{q-1}\cdot k^r,
\quad &\text{if}\ p=0, 1\ \&\ q\ge 1,
\\
2^{p-2}\cdot k^r,
\quad &\text{if}\  p\ge 2\ \&\ q=0,
\\ 
2^{p-2}\cdot 3^{q-1}\cdot k^r,
\quad &\text{if}\ p\ge 2\ \&\ q\ge 1.
 \end{array}
\right.
\end{align}
Correspondingly, we have 
\begin{align}
\label{cobordism_Zn}
\Omega_{\mathrm{Spin}}^{5}(B\mathbb{Z}_{n})=\Gamma_{\mathrm{Spin}}^5(B\mathbb{Z}_n) 
\cong\mathbb{Z}_{a_n} \times \mathbb{Z}_{b_n},
\quad\forall n\in\mathbb{N}.
\end{align}
  
In principle, one can use the combinatorial formula (\ref{combinatorial_eta}) to compute the values of the $\eta$ invariant on the associated generators to see the dependence of elements of $\Omega^{5}_{\mathrm{Spin}}(B\mathbb{Z}_n)$ on representations of $\mathbb{Z}_n$; however, working in this way is somehow not very useful (and systematic).
Instead of using the formula (\ref{combinatorial_eta}), one can compute the mod $\mathbb{Z}$ $\eta$ invariant, which is a bordism invariant, in terms of the more familiar A-roof polynomials that appear in the index theorem for Dirac operators. 
More specifically, we have (as shown in Appendix \ref{5d eta invariants from 7d})
\begin{align}
\label{eta_5d_7d_all_n}
\eta([X], R)=\eta(L(n; 1, 1, 1, 1), \tau_n([X])R)  \mod\mathbb{Z},
\quad \forall n\in\mathbb{N},
\end{align}
for any $[X]\in\Omega^{\mathrm{Spin}}_5(B\mathbb{Z}_n)$ and any representation $R\in RU(\mathbb{Z}_n)$.
Here $L(n; 1, 1, 1, 1) = S^7/ (\rho_{1}\oplus\rho_{1}\oplus\rho_{1}\oplus\rho_{1}$) is a seven-dimensional lens space and $\tau_n$ is some isomorphism from $\Omega^{\mathrm{Spin}}_5(B\mathbb{Z}_n)$ to $I_n/\{I_n\cap RU_0(\mathbb{Z}_n)^4\}$ (which exists by the relation (\ref{Gamma_iso_rep_theory})).
The mod $\mathbb{Z}$ $\eta$ invariant on $L(n; 1, 1, 1, 1)$ with the representation $\rho_s$ can be evaluated as \cite{Gilkey1984}
\begin{align}
\label{eta_A-roof}
\eta(L(n; 1, 1, 1, 1), \rho_s)
= -\frac{1}{n}\hat{A}_4 (s+n/2; n, 1, 1, 1, 1) \mod\mathbb{Z},
\end{align}
where
\begin{align}
\label{A-roof}
\hat{A}_k (t; \vec{x}) := \sum_{a+2b=k} t^a\hat{A}_b(\vec{x})/a!,
\end{align}
with $\hat{A}_k(\vec{x})$ being the A-roof polynomials.
The first few values of $\hat{A}_k(\vec{x})$ are
\begin{align}
\hat{A}_0(\vec{x})=1, 
\quad
\hat{A}_1(\vec{x})=-\frac{1}{24}\sum_i x_i^2, 
\quad
\hat{A}_2(\vec{x})=\frac{1}{5760}\left[-4\sum_{i<j}x_i^2 x_j^2 + 7(\sum_i x_i^2)^2  \right].
\quad
\end{align}

While $\tau_n$ is an isomorphism between $\Omega^{\mathrm{Spin}}_5(B\mathbb{Z}_n)$ and $I_n/\{I_n\cap RU_0(\mathbb{Z}_n)^4\}$, any set of generators of the former must be mapped to a set of generators of the latter. Recall that $I_n/\{I_n\cap RU_0(\mathbb{Z}_n)^4\}$ is generated by $\rho_1-\rho_{-1}$ and $\rho_2-\rho_{-2}$ with definite finite orders (the order of $\rho_1-\rho_{-1}$ modulo $RU_0(\mathbb{Z}_n)^4$ is $a_n$ defined in (\ref{def_an_bn})). Correspondingly, there exist two bordism classes $[X_n]$ and $[Y_n]$ as generators of $\Omega^{\mathrm{Spin}}_5(B\mathbb{Z}_n)$ 
such that 
\begin{align}
\label{eta_X_n}
\eta([X_n], \rho_{s}) 
&= \eta(L(n; 1, 1, 1, 1), (\rho_1-\rho_{-1})\rho_{s}) \mod\mathbb{Z}
\nonumber\\
&=- \frac{1}{n} \left[\hat{A}_4 (s+1+n/2; n, 1, 1, 1, 1) - \hat{A}_4 (s-1+n/2; n, 1, 1, 1, 1)\right] \mod\mathbb{Z}
\nonumber\\
& =-\frac{1}{6n} \left(2s^3+3ns^2+n^2s\right) \mod\mathbb{Z}
\nonumber\\
& =-\frac{1}{6n}\left(n^2+3n+2\right) s^3 \mod\mathbb{Z}.
\end{align}
\footnote{
For example, one can take $X_n = -X(n; 1, 1)$, the lens space bundle defined in (\ref{lens_space_bundele}). In fact, we have 
$
\eta(-X(n; 1, 1), R)=\eta(L(n; 1, 1, 1, 1), (\rho_1-\rho_{-1})R)
$
exactly.
}
and
\begin{align}
\label{eta_Y_n}
\eta([Y_n], \rho_{s}) 
&= \eta(L(n; 1, 1, 1, 1), (\rho_2-\rho_{-2})\rho_{s}) \mod\mathbb{Z}
\nonumber\\
&=- \frac{1}{n} \left[\hat{A}_4 (s+2+n/2; n, 1, 1, 1, 1) - \hat{A}_4 (s-2+n/2; n, 1, 1, 1, 1)\right] \mod\mathbb{Z}
\nonumber\\
& =-\frac{1}{3n}  \left[2s^3+(n^2+6)s \right] \mod\mathbb{Z}
\nonumber\\
&= 2 \eta([X_n], R) - \frac{2}{n}s \mod\mathbb{Z}.
\end{align}

Let $R$ be a generic (spin 1/2) $\mathbb{Z}_n$ representation including both left- and right-handed chiralities, that is, $R=R_L- R_R = \oplus_L \rho_{s_L} - \oplus_R \rho_{s_R}$. An element of $\Omega^{5}_{\mathrm{Spin}}(B\mathbb{Z}_n)$ associated with $R$ is represented by $\exp(-2\pi i\eta_R)$, characterizing a five-dimensional fermionic SPT phase or the anomaly of a four-dimensional Weyl fermions in the same representation. Here $\eta_R$ is the $\eta$ invariant map and can be represented by 
\begin{align}
(\eta([X_n], R),\ \eta([Y_n], R)),
\end{align}
or equivalently by 
\begin{align}
\label{anomaly_untwisted}
\alpha_R&:= (\eta([-X_n], R),\ \eta([2X_n-Y_n], R)) 
\nonumber\\
&= \left(\frac{1}{6n}\left(n^2+3n+2\right) \Delta s_3\mod\mathbb{Z}, 
\quad \frac{2}{n}\Delta s_1 \mod\mathbb{Z} \right),
\end{align}
\footnote{
We have to clarify that, in this expression, the first element is a generator of the $\mathbb{Z}_{a_n}$ group in (\ref{cobordism_Zn}), while the second element is in general \emph{not} a generator of the $\mathbb{Z}_{b_n}$ group in (\ref{cobordism_Zn}).
}
where
$\Delta s_3 := \sum_{L} s_L^3 - \sum_Rs_R^3$
and $\Delta s_1 :=\sum_{L} s_L - \sum_Rs_R$.
Note that the second term $\frac{2}{n}\Delta s_1 \mod\mathbb{Z}$ in ($\ref{anomaly_untwisted}$) corresponds to the mixed 't Hooft anomaly of $\mathrm{Spin}(4)\times\mathbb{Z}_n$, 
\footnote{
Such a mixed anomaly can also be represented by the 5d cobordism class $p_1\nu$, where $p_1$ is the first Pontryagin class and $\nu\in H^1(B\mathbb{Z}_n, U(1))\cong\mathbb{Z}_n$. 
}
while $\alpha_R$ is the 't Hooft anomaly of the whole $\mathrm{Spin}(4)\times\mathbb{Z}_n$ group.
One may also regard the first term in ($\ref{anomaly_untwisted}$) as the ``purely gauge anomaly'' and the second term ($\ref{anomaly_untwisted}$) as ``mixed gauge-gravitational anomaly'' --- like what people do in the case of U(1) chiral gauge theory.

A trivial $\alpha_R$ corresponds to an anomaly-free representation, where the mod $n$ charges satisfy
\begin{align}
\label{anomaly-free_Z_n}
\left(n^2+3n+2\right) \Delta s_3&= 0 \mod 6n, 
\quad 2\Delta s_1=0 \mod n,
\nonumber\\
\text{or}
\quad\Delta s_3  &= 0 \mod a_n,
\quad\Delta s_1 = 0 \mod n/2,
\end{align}
with $a_n$ given in (\ref{def_an_bn}).




\subsection{'t Hooft anomalies of $\mathrm{Spin}^{\mathbb{Z}_{2m}}(4)$}
\label{'t Hooft anomalies of Spin(4)^Z2m}

Now let us consider Weyl fermions with a twisted $\mathbb{Z}_{2m}$ symmetry, that is, fermions transforming under the $\mathrm{Spin}^{\mathbb{Z}_{2m}}(4)$ group. We would like to determine the group $\Gamma^{5}_{\mathrm{Spin}^{\mathbb{Z}_{2m}}}$, a subgroup of 
\begin{align}
\Omega^{5}_{\mathrm{Spin}^{\mathbb{Z}_{2m}}} :=
\mathrm{Hom}(\Omega_{5}^{\mathrm{Spin}^{\mathbb{Z}_{2m}}}, \mathrm{U}(1))
=\mathrm{Hom}(\Omega_{5}^{\mathrm{Spin}}(B\mathbb{Z}_m, \xi), \mathrm{U}(1))
\end{align}
that is generated by the (exponentiated) $\eta$ invariants (that is, free fermions), as well as 
the anomaly $\tilde{\alpha}_{\tilde{R}}$ (defined later) of $\Psi$ in an arbitrary $\mathrm{spin}^{\mathbb{Z}_{2m}}$ representation $\tilde{R}$. Note that all the mod $2m$ charges $\tilde{s}$ in $\tilde{R}$ must be odd integers, that is,
\begin{align}
\tilde{R}\in RU^o(\mathbb{Z}_{2m}):=\oplus_{\tilde{s}\in\mathrm{odd}}\ \tilde{\rho}_{\tilde{s}}\cdot\mathbb{Z},
\quad\tilde{\rho}_{\tilde{s}}=e^{\pi i\tilde{s}/m}.
\end{align}
While our discussion also involves representations of $\mathbb{Z}_m=\mathbb{Z}_{2m}/\mathbb{Z}_2$, we use the untilded $R$ to denote an element of $RU(\mathbb{Z}_{m})$ to avoid confusion. 
Finally, we denote $\eta(X, \tilde{R})$ as the $\eta$ invariant on a closed five-dimensional $\mathrm{Spin}^{\mathbb{Z}_{2m}}$ manifold $X$ throughout this section.

Within a similar discussion as in the case of untwisted $\mathbb{Z}_n$ symmetries, we first have to compute $\Gamma_{5}^{\mathrm{Spin}^{\mathbb{Z}_{2m}}}$, which is the Pontryagin dual of $\Gamma^{5}_{\mathrm{Spin}^{\mathbb{Z}_{2m}}}$. The equivalence relation among elements of $\Gamma_{5}^{\mathrm{Spin}^{\mathbb{Z}_{2m}}}$ is defined in an analogous fashion: 
If $X_1$ and $X_2$ are two five-dimensional spin$^{\mathbb{Z}_{2m}}$ manifolds, $X_1\sim X_2$ if $\eta(X_1 - X_2, \tilde{R})=0 \mod \mathbb{Z}$ for all spin$^{\mathbb{Z}_{2m}}$ representations $\tilde{R}\in RU^o(\mathbb{Z}_{2m})$, and we denote $[X]_{\eta}\in\Gamma_{5}^{\mathrm{Spin}^{\mathbb{Z}_{2m}}}$ be the equivalence class of $X$.

As shown in \cite{Botvinnik:1997aa, Barrera-Yanez:2006aa}, $\Gamma_{5}^{\mathrm{Spin}^{\mathbb{Z}_{2m}}}$ is generated by the five-dimensional lens spaces $L(m; a_1, a_2, a_3) := S^5/(\rho_{a_1}\oplus\rho_{a_2}\oplus\rho_{a_3})$ --- which are spin$^{\mathbb{Z}_{2m}}$ (and spin$^G$) manifolds --- with various spin$^{\mathbb{Z}_{2m}}$ structures. Furthermore, there is a combinatorial formula for the $\eta$ invariant on a generic ($4j+1$)-dimensional lens space $L(m; \vec{a})$ with the spin$^{\mathbb{Z}_{2m}}$ structure given by a real 2-plane bundle $\xi$ defined by $\rho_1$ over $B\mathbb{Z}_m$:
\footnote{
$L(m; \vec{a})$ admits a natural spin$^G$ structure with determinant line bundle given by $\rho_1$, and $\xi$ is the underlying real 2-plane bundle of this complex line bundle.
}
\begin{align}
\label{combinatorial_eta_lens_space_spinZ2m}
\eta(L(m; a_1, a_2, ..., a_{2j+1}), R) =  \frac{1}{m} \sum_{\lambda\in\mathbb{Z}_m, \lambda\neq 1} \mathrm{Tr}(R(\lambda))
\frac{\lambda^{\frac{1}{2}(a_1+\ \cdots\ +a_{2j+1}+1)}}{(1-\lambda^{a_1})(1-\lambda^{a_2})\cdots(1-\lambda^{a_{2j+1}})},
\end{align}
where $R=\oplus_sk_s\rho_s\in\ RU(\mathbb{Z}_m)$. The above formula can also be expressed in terms of a spin$^{\mathbb{Z}_{2m}}$ representation $\tilde{R}=\oplus_{\tilde{s}}\tilde{k}_{\tilde{s}}\tilde{\rho}_{\tilde{s}}\in RU^o(\mathbb{Z}_{2m})$, via the relation 
\begin{align}
\label{spinZ2m_rep_to_Zm_rep}
\tilde{s} = 2s+1 \mod 2m, \quad \tilde{k}_{\tilde{s}}=k_s \in\mathbb{Z}.
\end{align}
Like the case of untwisted $\mathbb{Z}_n$ symmetries, one can also identify $\Gamma_{5}^{\mathrm{Spin}^{\mathbb{Z}_{2m}}}$ in terms of the representation theory of $\mathbb{Z}_m$. Specifically, 
\begin{align}
\label{Gamma_iso_rep_theory_twisted}
\Gamma_{5}^{\mathrm{Spin}^{\mathbb{Z}_{2m}}} \cong \tilde{I}_m/\{\tilde{I}_m\cap RU_0(\mathbb{Z}_m)^6\},
\quad\forall m\in\mathbb{N},
\end{align}
where 
\begin{align}
\tilde{I}_{m} := \oplus_{j\ge1}(\rho_0-\rho_1)^{2j}\rho_{-1}^j\cdot\mathbb{Z}.
\end{align}
This can be verified, following a similar approach in Sec.~\ref{'t Hooft anomalies of Spin(4)xZn} and Appendix~\ref{isomorphisms}, by relating the $\eta$ invariants (\ref{combinatorial_eta_lens_space_spinZ2m}) on some specific 5d ($j=1$) lens spaces to those on 9d ($j=2$) lens spaces to construct the above isomorphism for each $m$. 
A detailed discussion on the case $m=2^v$ can be found in \cite{Barrera-Yanez:2006aa}.
\footnote{
In fact, the author of Ref. \cite{Barrera-Yanez:2006aa} showed that
$
\Gamma_{5}^{\mathrm{Spin}^{\mathbb{Z}_{2m}}} \cong \tilde{I}_m/\{\tilde{I}_m\cap RU_0(\mathbb{Z}_m)^5\}
$, with the power of $RU_0(\mathbb{Z}_m)$ different from the one appearing in (\ref{Gamma_iso_rep_theory_twisted}). However, by carefully going through the proofs in \cite{Barrera-Yanez:2006aa} as well as by the computation result in this paper, we confirmed that (\ref{Gamma_iso_rep_theory_twisted}) is the correct one. 
}

The representation theory $\tilde{I}_m/\{\tilde{I}_m\cap RU_0(\mathbb{Z}_m)^6\}$ can be expressed in terms of  cyclic groups. Expanding
\begin{align}
&\tilde{I}_m/\{\tilde{I}_m\cap RU_0(\mathbb{Z}_m)^6\}
\nonumber\\
&=\{(\rho_0-\rho_{1})^2\rho_{-1}\cdot\mathbb{Z}\oplus(\rho_0-\rho_{1})^4\rho^2_{-1}\cdot\mathbb{Z}\}
/\tilde{I}_m\cap RU_0(\mathbb{Z}_m)^6\}
\nonumber\\
&=
\{(\rho_0-\rho_{1})^2\rho_{-1}\cdot\mathbb{Z}\oplus(\rho_0-\rho_{2})^2\rho_{-2}\cdot\mathbb{Z}\}
/\tilde{I}_m\cap RU_0(\mathbb{Z}_m)^6\}
\nonumber\\
&=
\{\{(1+x)+(1+x)^{m-1}-2\}\cdot\mathbb{Z}\oplus\{(1+x)^2+(1+x)^{m-2}-2\}\cdot\mathbb{Z}\}
\nonumber\\
&\ \quad/\{\{(1+x)^m-1\}\cdot\mathbb{Z}[x]+x^6\cdot\mathbb{Z}[x]\}
\nonumber\\
&=
\left\{
\left\{mx
+\left(
 \begin{smallmatrix}
m-1\\ 2
\end{smallmatrix}
\right) 
x^2+
\left(
 \begin{smallmatrix}
m-1\\ 3
\end{smallmatrix}
\right) 
x^3+
\left(
 \begin{smallmatrix}
m-1\\ 4
\end{smallmatrix}
\right) 
x^4+
\left(
 \begin{smallmatrix}
m-1\\ 5
\end{smallmatrix}
\right) 
x^5
\right\}\cdot\mathbb{Z}
\right.
\nonumber\\
&\ \quad\left.\oplus
\left\{mx
+\left[\left(
 \begin{smallmatrix}
m-2\\ 2
\end{smallmatrix}
\right) +1\right]
x^2+
\left(
 \begin{smallmatrix}
m-2\\ 3
\end{smallmatrix}
\right) 
x^3+
\left(
 \begin{smallmatrix}
m-2\\ 4
\end{smallmatrix}
\right) 
x^4+
\left(
 \begin{smallmatrix}
m-2\\ 5
\end{smallmatrix}
\right) 
x^5
\right\}\cdot\mathbb{Z}
\right\}
\nonumber\\
&\quad\
/\left\{
\left\{mx
+\left(
 \begin{smallmatrix}
m\\ 2
\end{smallmatrix}
\right) 
x^2+
\left(
 \begin{smallmatrix}
m\\ 3
\end{smallmatrix}
\right) 
x^3+
\left(
 \begin{smallmatrix}
m\\ 4
\end{smallmatrix}
\right) 
x^4+
\left(
 \begin{smallmatrix}
m\\ 5
\end{smallmatrix}
\right) 
x^5
\right\}\cdot\mathbb{Z}[x]
+
x^6\cdot\mathbb{Z}[x]
\right\},
\end{align}
we found
\begin{align}
\label{iso_Im_cyclic}
\tilde{I}_m/\{\tilde{I}_m\cap RU_0(\mathbb{Z}_m)^6\}
&\cong
\left\{
\begin{array}{ll}
0,
\quad &\text{if}\ m=1,
 \\
\mathbb{Z}_{8m} \oplus \mathbb{Z}_{m/2},
\quad &\text{if}\ m=2^v>1,
\\
\mathbb{Z}_{3m} \oplus \mathbb{Z}_{m/3},
\quad &\text{if}\ m=3^v,
\\ 
\mathbb{Z}_{m} \oplus \mathbb{Z}_{m},
\quad &\text{if}\ m=p^v,\ p>3,
 \end{array}
\right.
\end{align}
for $m$ equal to a prime power, and in general 
\begin{align}
\label{iso_Im_cyclic_2}
\tilde{I}_m/\{\tilde{I}_m\cap RU_0(\mathbb{Z}_m)^6\}
&\cong\mathbb{Z}_{\tilde{a}_m}\oplus\mathbb{Z}_{\tilde{b}_m}.
\end{align}
Here the integers $\tilde{a}_m$ and $\tilde{b}_m$ are defined, when expressing $m=2^{p}\cdot 3^{q}\cdot k^{r}$ with $k\ge 5$ being an odd number not divisible by 3 and $p, q, r$ being nonnegative integers, as
\begin{align}
\label{def_am_bm}
\tilde{a}_m &:=
\left\{
\begin{array}{ll}
k^r,
\quad &\text{if}\ p=0, 1\ \&\ q=0,
 \\
3^{q+1}\cdot k^r,
\quad &\text{if}\ p=0, 1\ \&\ q\ge 1,
\\
2^{p+3}\cdot k^r,
\quad &\text{if}\  p\ge 1\ \&\ q=0,
\\ 
2^{p+3}\cdot 3^{q+1}\cdot k^r,
\quad &\text{if}\ p\ge 1\ \&\ q\ge 1,
 \end{array}
\right.
\nonumber\\
\tilde{b}_m &:=
\left\{
\begin{array}{ll}
k^r,
\quad &\text{if}\ p=0, 1\ \&\ q=0,
 \\
3^{q-1}\cdot k^r,
\quad &\text{if}\ p=0, 1\ \&\ q\ge 1,
\\
2^{p-1}\cdot k^r,
\quad &\text{if}\  p\ge 1\ \&\ q=0,
\\ 
2^{p-1}\cdot 3^{q-1}\cdot k^r,
\quad &\text{if}\ p\ge 1\ \&\ q\ge 1.
 \end{array}
\right.
\end{align}

The spin$^{\mathbb{Z}_{2m}}$ bordism group $\Omega_{5}^{\mathrm{Spin}^{\mathbb{Z}_{2m}}}= \Omega_{5}^{\mathrm{Spin}}(B\mathbb{Z}_m, \xi)$ for each $m\in \mathbb{N}$  is identical to $\Gamma_{5}^{\mathrm{Spin}^{\mathbb{Z}_{2m}}}$, with the same expression of direct sums of cyclic groups in (\ref{iso_Im_cyclic_2}). This can be confirmed again by noting that the upper bound of the order of $\Omega_{5}^{\mathrm{Spin}^{\mathbb{Z}_{2m}}}$ set by the Atiyah-Hirzebruch spectral sequence for each $m=2^{p}\cdot \ell^{q}$ with an odd number $\ell$, that is, $2^{2p+2}\cdot\ell^{2q}$ 
\footnote{
Specifically, for $m=2^{p}\cdot \ell^{q}$ we have \cite{Gilkey, Tachikawa:2018aa}
\begin{align}
&|\Omega^{\mathrm{Spin}}_{5}(B\mathbb{Z}_{m}, \xi)|
\le \prod_{a+b=5}|\tilde{H}_a(B\mathbb{Z}_m, \Omega^{\mathrm{Spin}}_b(pt))|
\nonumber\\
&= |\tilde{H}_0(B\mathbb{Z}_m, 0)| \cdot |\tilde{H}_1(B\mathbb{Z}_m, \mathbb{Z})| \cdot |\tilde{H}_2(B\mathbb{Z}_m, 0)|
\cdot |\tilde{H}_3(B\mathbb{Z}_m, \mathbb{Z}_2)| \cdot |\tilde{H}_4(B\mathbb{Z}_m, \mathbb{Z}_2)| \cdot |\tilde{H}_5(B\mathbb{Z}_m, \mathbb{Z})|
\nonumber\\
&=2^{2p+2}\cdot\ell^{2q} 
\nonumber
\end{align}
},
matches the order of $\Gamma_{5}^{\mathrm{Spin}^{\mathbb{Z}_{2m}}}$. 
In particular, the result $\Omega_{5}^{\mathrm{Spin}^{\mathbb{Z}_{4}}}\cong\mathbb{Z}_{16}$ agrees with the one computed in \cite{Tachikawa:2018aa}.
The cobordism groups can also be determined accordingly:
\begin{align}
\label{cobordism_Z2m}
\Omega_{\mathrm{Spin}^{\mathbb{Z}_{2m}}}^{5}=\Gamma_{\mathrm{Spin}^{\mathbb{Z}_{2m}}}^{5}
\cong\mathbb{Z}_{\tilde{a}_m}\times\mathbb{Z}_{\tilde{b}_m},
\quad\forall m\in\mathbb{N}.
\end{align}

Finally, we would like to know the dependence of elements of $\Omega_{\mathrm{Spin}^{\mathbb{Z}_{2m}}}^{5}$ on spin$^{\mathbb{Z}_{2m}}$ representations. Instead of directly computing the the $\eta$ invariants on generators of $\Omega^{\mathrm{Spin}^{\mathbb{Z}_{2m}}}_{5}$, we use a similar technique as we did in the untwisted case to perform the computation. According to the relation (\ref{Gamma_iso_rep_theory_twisted}), the spin$^{\mathbb{Z}_{2m}}$ $\eta$ invariant on 5d lens spaces can be identified by the one on a 9d lens space, through the following relation:
\begin{align}
\label{eta_5d_9d_all_m}
\eta([X], R)=\eta(L(m; 1, 1, 1, 1,1), \tilde{\tau}_m([X])R)  \mod\mathbb{Z},
\quad \forall m\in\mathbb{N},
\end{align}
for any $[X]\in\Omega^{\mathrm{Spin}^{\mathbb{Z}_{2m}}}_{5}$ and $R\in\ RU(\mathbb{Z}_m)$ (associated to $\tilde{R}\in RU^o(\mathbb{Z}_{2m})$ by (\ref{spinZ2m_rep_to_Zm_rep})), where $\tilde{\tau}_m$ is an isomorphism between $\Omega^{\mathrm{Spin}^{\mathbb{Z}_{2m}}}_{5}$ and $\tilde{I}_m/\{\tilde{I}_m\cap RU_0(\mathbb{Z}_m)^6\}$. Since the latter is generated by  $(\rho_0-\rho_{1})^2\rho_{-1}$ and $(\rho_0-\rho_{1})^4\rho^2_{-1}$ (modulo $RU_0(\mathbb{Z}_m)^6$), we have a corresponding set of generators $\{[\tilde{X}]_m, [\tilde{Y}]_m\}$ of $\Omega^{\mathrm{Spin}^{\mathbb{Z}_{2m}}}_{5}$ with the values of $\eta$ invariant as 
\begin{align}
\label{eta_X_m_Y_m}
\eta([\tilde{X}]_m], R) 
&= \eta(L(m; 1, 1, 1, 1,1), (\rho_0-\rho_{1})^2\rho_{-1}\cdot R) \mod\mathbb{Z}
\nonumber\\
\eta([\tilde{Y}]_m], R) 
&= \eta(L(m; 1, 1, 1, 1,1), (\rho_0-\rho_{1})^4\rho^2_{-1}\cdot R) \mod\mathbb{Z}.
\end{align}
\footnote{
For example, one can take $X_m$ as the 5d lens space $L(m, 1,1,1)$:
$
\eta(L(m, 1,1,1), R)=\eta(L(m; 1, 1, 1, 1,1), (\rho_0-\rho_{1})^2\rho_{-1}\cdot\rho_{s}R).
$
}

On the other hand, we can also relate the $\eta$ invariant on a 9d lens space with the spin$^{\mathbb{Z}_{2m}}$ structure to the $\eta$ invariant on a 7d lens space with the spin$\times\mathbb{Z}_m$ structure through the the two formulas (\ref{combinatorial_eta_lens_space_spinZ2m}) and (\ref{combinatorial_eta_lens_space_spinxZn}):
\begin{align}
\label{eta_7d_9d}
\eta_{\mathrm{spin}^{\mathbb{Z}_{2m}}}(L(m; 1, 1, 1, 1,1), (\rho_0-\rho_{1})^2\rho_{-1}\cdot\rho_{s}) 
&= \eta_{\mathrm{spin}\times\mathbb{Z}_m}(L(m; 1, 1, 1, 1), (\rho_0-\rho_{1})\rho_{s}) 
\nonumber\\
\eta_{\mathrm{spin}^{\mathbb{Z}_{2m}}}(L(m; 1, 1, 1, 1,1), (\rho_0-\rho_{1})^4\rho_{-1}^2\cdot\rho_{s}) 
&= \eta_{\mathrm{spin}\times\mathbb{Z}_m}(L(m; 1, 1, 1, 1), (\rho_0-\rho_{1})^3\rho^{-1}\cdot\rho_{s}),
\end{align}
Then, using the expression (\ref{eta_A-roof}) as well as the relation (\ref{spinZ2m_rep_to_Zm_rep}), we obtain
\begin{align}
\label{eta_X_m_2}
&\eta([\tilde{X}]_m], \tilde{\rho}_{\tilde{s}}) 
\nonumber\\
&=- \frac{1}{m} \left[\hat{A}_4 ((\tilde{s}-1)/2+m/2; m, 1, 1, 1, 1) - \hat{A}_4 ((\tilde{s}-1)/2+1+m/2; m, 1, 1, 1, 1)\right] \mod\mathbb{Z}
\nonumber\\
& =\frac{1}{48m} \left[\tilde{s}^3+3m\tilde{s}^2+(2m^2-3)\tilde{s}-3m\right] \mod\mathbb{Z}
\nonumber\\
& =\frac{1}{48m} \left[(2m^2+m+1)\tilde{s}^3-(m+3)\tilde{s}\right] \mod\mathbb{Z}
\end{align}
and
\begin{align}
\label{eta_Y_m_2}
&\eta([\tilde{Y}]_m], \tilde{\rho}_{\tilde{s}}) 
\nonumber\\
&=- \frac{1}{m} \left[\hat{A}_4 ((\tilde{s}-1)/2-1+m/2; m, 1, 1, 1, 1) -3 \hat{A}_4 ((\tilde{s}-1)/2+m/2; m, 1, 1, 1, 1)\right. 
\nonumber\\
&\ \ \quad\qquad\left.+3\hat{A}_4 ((\tilde{s}-1)/2+1+m/2; m, 1, 1, 1, 1) - \hat{A}_4 ((\tilde{s}-1)/2+2+m/2; m, 1, 1, 1, 1)\right] \mod\mathbb{Z}
\nonumber\\
& =\frac{1}{2m} (\tilde{s}+m) \mod\mathbb{Z}
\nonumber\\
& =\frac{1}{2m} (\tilde{s}+m\tilde{s}^3) \mod\mathbb{Z}.
\end{align}

Therefore, an element of $\Omega_{\mathrm{Spin}^{\mathbb{Z}_{2m}}}^{5}$ associated with a generic spin$^{\mathbb{Z}_{2m}}$ representation 
$\tilde{R}=\tilde{R}_L-\tilde{R}_R = \oplus_L \tilde{\rho}_{\tilde{s}_L} - \oplus_R \tilde{\rho}_{\tilde{s}_R}$ corresponds to  $\exp(-2\pi i\eta_{\tilde{R}})$, with $\eta_{\tilde{R}}$ represented by
\begin{align}
\label{anomaly_twisted}
\tilde{\alpha}_{\tilde{R}}&:=(\eta([\tilde{X}_m], \tilde{R}),\ \eta([\tilde{Y}_m], \tilde{R}))
\nonumber\\
&= \left(\frac{1}{48m} \left[(2m^2+m+1)\Delta\tilde{s}_3-(m+3)\Delta\tilde{s}_1\right]\mod\mathbb{Z}, 
\quad \frac{1}{2m}\left(m\Delta\tilde{s}_3+\Delta \tilde{s}_1\right) \mod\mathbb{Z} \right),
\end{align}
\footnote{
In this expression, the first element is a generator of the $\mathbb{Z}_{\tilde{a}_m}$ group in (\ref{cobordism_Z2m}), while the second element is in general \emph{not} a generator of the $\mathbb{Z}_{\tilde{b}_m}$ group in (\ref{cobordism_Z2m}).
}
where
$\Delta \tilde{s}_3 := \sum_{L} \tilde{s}_L^3 - \sum_R\tilde{s}_R^3$
and $\Delta \tilde{s}_1 :=\sum_{L} \tilde{s}_L - \sum_R\tilde{s}_R$.
Note that all the spin$^{\mathbb{Z}_{2m}}$ charges $\tilde{s}_{L,R}$ must be odd number modulo $2m$. 
Observe that in the form of $\tilde{\alpha}_{\tilde{R}}$ the cubic term $\Delta\tilde{s}_3$ and the linear term $\Delta\tilde{s}_1$ couple to each other. So one can not really distinguish which one as the ``purely gauge anomaly'' and which one as ``mixed gauge-gravitational anomaly'' --- as we are considering the $\mathrm{Spin}^{\mathbb{Z}_{2m}}(4)$ group where the spacetime symmetry is twisted with the internal symmetry.

Any anomaly-free representation $\tilde{R}$ corresponds to a trivial $\tilde{\alpha}_{\tilde{R}}$, that is,
\begin{align}
\label{anomaly-free_twisted}
(2m^2+m+1)\Delta\tilde{s}_3-(m+3)\Delta\tilde{s}_1= 0 \mod 48m,
\quad m\Delta\tilde{s}_3+\Delta \tilde{s}_1=0 \mod 2m.
\end{align}


\section{Further discussion of anomalies of 4d chiral gauge theory associated with discrete symmetries}
\label{Further discussion of anomalies of 4d chiral gauge theory associated with discrete symmetries}

In the last section we computed the 't Hooft anomalies of the $\mathrm{Spin}(4)\times\mathbb{Z}_n$ and the $\mathrm{Spin}(4)^{\mathbb{Z}_{2m}}$ groups. The anomaly-free condition (\ref{anomaly-free_Z_n})/(\ref{anomaly-free_twisted}) tell us when we can consistently couple a set of chiral fermions to a $\mathbb{Z}_n$/$\mathrm{spin}^{\mathbb{Z}_{2m}}$ gauge field, which can be either classical or dynamical.
In this section, we compare these conditions to those obtained by Ib\'{a}\~{n}ez and Ross, which we review briefly as follows. 


\subsection{Review of the Ib\'{a}\~{n}ez-Ross conditions for discrete symmetry anomalies}
\label{Review of the Ib\'{a}\~{n}ez-Ross conditions for discrete symmetry anomalies}

Before stating their argument, we would like to mention that the result obtained by Ib\'{a}\~{n}ez and Ross in \cite{Ibanez-Ross1991} is only for the case of untwisted $\mathbb{Z}_n$ symmetries, but their approach can also be applied to the twisted cases without without any difficulty.

Ib\'{a}\~{n}ez and Ross argued the anomaly constraints on $\mathbb{Z}_n$ charges of a set of massless Weyl  fermions by embedding the untwisted $\mathbb{Z}_n$ gauge symmetry in a U(1) gauge symmetry. These constraints are derived basing on the anomaly cancellation conditions of the U(1) symmetry, which involve both the (perturbative) gauge and mixed gauge-gravitational anomalies, together with the constraints on the charges of the fermions that acquire mass through spontaneous breaking  of U(1).
Specifically, let $\{\{q_i\}, \{Q_j\} \}$ be the U(1) charges of a collection of left-handed Weyl fermions.
\footnote{
The contribution of a left-handed Weyl fermion of charge $q$ to the anomaly is equal to a right-handed Weyl fermon of charge $-q$. Without loss of generality, one can just consider fermions with a specific chirality to derive the anomaly constraints.
}
To guarantee that the theory is anomaly-free, these charges must obey the relations
$\sum_i q_i^3 + \sum_j Q_j^3 = 0$ and 
$\sum_i q_i + \sum_j Q_j = 0$.
We then introduce a Higgs field $\phi$ of charge $n$ to spontaneously break the U(1) symmetry down to a $\mathbb{Z}_n$ symmetry, and also add Yukawa couplings between the $\phi$ field and the charge-$Q_j$ fermions, so that these fermions gain mass from the expectation value of $\phi$ (while the charge-$q_i$ fermions are left massless in the low energy phase). 
A generic Yukawa coupling includes the Dirac-type mass terms, which couple each pair of different Weyl fermions, and the Majorana-type mass terms, which couple each Weyl fermion with itself.
As these mass terms are required to be gauge invariant when coupled to single-valued potentials of the Higgs field, the charges of the massive fermions must obey $Q_{j'} + Q_{j''} = \mathrm{integer}\times n$ for each pair of fermions with a Dirac mass and, if $n$ is even, $2Q_{l} = \mathrm{integer}\times n$ for each fermion with a Majorana mass. Then, writing $q_i = s_i+m_i n$, where $s_i, m_i\in\mathbb{Z}$ and $0\leq s_i<n$, the U(1) anomaly cancellation conditions plus the charge constraints on the massive states yield
\begin{align}
\label{IR_condition_untwisted}
&\qquad\sum_i s_i^3 = pn+ r\frac{n^3}{8},\quad \sum_i s_i = p'n+ r'\frac{n}{2},
 \nonumber\\
&p, r, p', r'\in\mathbb{Z};\ p \in 3\mathbb{Z}\ \text{if}\ n\in 3\mathbb{Z},\ r, r' = 0\ \text{if $n$ is odd}.
\end{align}

The above equation is the so-called \emph{Ib\'{a}\~{n}ez-Ross condition} for an anomaly-free $\mathbb{Z}_n$ gauge symmetry of massless fermions.
It is understood to be a necessary but not sufficient condition --- as the $\mathbb{Z}_n$ gauge theory  is assumed to be embedded in some U(1) gauge theory. 
In particular, for a given set of $\mathbb{Z}_n$ charges $\{s_i\}$ and any given integers $\{p, r, p', r'\}$ that (\ref{IR_condition_untwisted}) is satisfied, we are even not sure if there always exists a high energy U(1) gauge theory in which the $\mathbb{Z}_n$ gauge theory can be embedded.
Also note that, in deriving the condition~(\ref{IR_condition_untwisted}), we have implicitly assumed that all the U(1) charges have integer values and massive fermions (after U(1) is broken) of integer charges do not contribute to cancellation of the anomaly of a low energy $\mathbb{Z}_n$ gauge group.

 The constraint in (\ref{IR_condition_untwisted}) that is linear in the $\mathbb{Z}_n$ charges can also be argued by considering the violation of the low energy $\mathbb{Z}_n$ symmetry in the presence of a gravitational instanton which is a spin manifold
 \cite{Banks-Dine1991, Araki2007},
 \footnote{
 One can also constrain the $\mathbb{Z}_n$ symmetry by introducing gauge instantons of a continuous gauge symmetry to the theory, as argued in \cite{PTWW1991, Banks-Dine1991}. Here we consider the situation that only gravitational instantons are present.
 }
without referring to information of any high energy theories in which the massless fermions are embedded.
On the other hand, the nonlinear (cubic) constraint, as pointed out by Banks and Dine in \cite{Banks-Dine1991},
 might be too restrictive and might not be required for consistency of the low energy theory, because it is not solely from the low energy considerations and would depend on assumptions about high energy theories. In particular, changes of the normalization of U(1) charges would affect this constraint. The cubic constraint could be weaker if we are not restricted to integer normalization of charges. It is always possible to make the $\mathbb{Z}_n$ charges of massless fermions satisfying the cubic constraint by extending the underlying $\mathbb{Z}_n$ symmetry to, for example, a $\mathbb{Z}_{n^2}$ symmetry (so that the massless particles transform under an effective $\mathbb{Z}_n$ symmetry while the whole theory --- including the massive degrees of freedom --- respects the  true $\mathbb{Z}_{n^2}$ symmetry).
 However, one can not do so for the linear constraint by modifying the massive content of the low energy theory, regardless of the normalization of the charges.
 
From this aspect, the linear constraint in (\ref{IR_condition_untwisted}) should be more respected than the nonlinear one in constraining the spectrum of light particles
 \cite{Banks-Dine1991}.
Failure of the nonlinear constraint implies only the existences of some ``fractionally'' charged massive states and an enlarged symmetry group at high energy. 
(Another point of view is that these fractionally charged states are indeed anomalous and contribute to cancellation of the anomaly of the low energy states
\cite{Ibanez1992}; however, it was not known at that time whether one can present the nonlinear constraint in a way that it throws much light on the nature of these states.)

Following the argument by Ib\'{a}\~{n}ez and Ross, we can get a similar anomaly cancellation condition for twisted $\mathbb{Z}_{2m}$ or $\mathrm{Spin}^{\mathbb{Z}_{2m}}(4)$ symmetries. Note that in this case all the massless fermions must have odd charges modulo $2m$. We consider cases with $m>1$ in the following, as  $\mathrm{Spin}^{\mathbb{Z}_{2}}(4)=\mathrm{Spin}(4)$ is trivial.
This time we can embed the underlying $\mathrm{spin}^{\mathbb{Z}_{2m}}$ gauge symmetry in a $\mathrm{spin}^c$ gauge symmetry where the U(1) charges of fermions are odd integers. Then, the anomaly constraints on $\mathrm{spin}^c$ charges together with the restriction of charges of the massive fermions coming from the Higgs mechanism gives ($m>1$)
\begin{align}
\label{IR_condition_twisted}
\sum_i \tilde{s}_i^3 = 2pm,\quad \sum_i \tilde{s}_i = 2p'm,
\quad p, p'\in\mathbb{Z};\ p \in 3\mathbb{Z}\ \text{if}\ n\in 3\mathbb{Z},
\end{align}
where $\tilde{s}_i$ are $\mathrm{spin}^{\mathbb{Z}_{2m}}$ charges of the underlying left-handed Weyl fermions. Note that on the right-hand side of above expression there is only contribution from the Dirac-type masses (as the Majorana-type masses are not allowed for fermions with odd U(1) charges). 
Like the case of untwisted $\mathbb{Z}_n$ symmetries, (\ref{IR_condition_twisted}) is just a necessary condition for an anomaly-free $\mathrm{spin}^{\mathbb{Z}_{2m}}$ gauge theory.

There is also an issue about the relation between anomaly constraints and symmetry extensions here. The condition (\ref{IR_condition_twisted}) must be satisfied if the total symmetry group of the low energy theory, including both the massless and massive (topological) parts, is specified as $\mathrm{Spin}^{\mathbb{Z}_{2m}}(4)$. However, in the case that the full symmetry is unknown --- while just having the massless sector transforming under an effective $\mathrm{Spin}^{\mathbb{Z}_{2m}}(4)$ symmetry --- it is possible to weaken the condition (\ref{IR_condition_twisted}) by enlarging the symmetry (in particular, on the massive sector). In the present situation, not only the cubic constraint but also the linear one in (\ref{IR_condition_twisted}) can change under a symmetry extension, e.g., from $\mathrm{Spin}^{\mathbb{Z}_{2m}}(4)$ to $\mathrm{Spin}(4)\times\mathbb{Z}_{2lm}$ for some $l\in\mathbb{N}$;  a new anomaly cancellation condition is arrived by the embedding of the low energy theory in a Spin(4)$\times$U(1) (rather than a $\mathrm{Spin}^{c}(4)$) chiral gauge theory, which has the form of (\ref{IR_condition_untwisted}).

\subsection{Connection to the Ib\'{a}\~{n}ez-Ross conditions}
\label{Connection to the Ib\'{a}\~{n}ez-Ross conditions}

The anomaly-free conditions (\ref{anomaly-free_Z_n}) and (\ref{anomaly-free_twisted}) we derived basing on the Dai-Freed theorem as well as on the topological classification of five-dimensional manifolds with the associated structures have similar forms as the the Ib\'{a}\~{n}ez-Ross conditions (\ref{IR_condition_untwisted}) and (\ref{IR_condition_twisted}) obtained by embedding the $\mathbb{Z}_n$/$\mathrm{spin}^{\mathbb{Z}_{2m}}$ gauge theories (coupled to Weyl fermions) in U(1)/spin$^c$ gauge theories. These equations involve only the linear terms and the cubic terms of the $\mathbb{Z}_n$/$\mathrm{spin}^{\mathbb{Z}_{2m}}$ charges.
However, (\ref{anomaly-free_Z_n})/(\ref{anomaly-free_twisted}) should be a necessary and sufficient condition for consistently gauging a $\mathbb{Z}_n$/$\mathrm{spin}^{\mathbb{Z}_{2m}}$ symmetry of a chiral fermion theory, while (\ref{IR_condition_untwisted})/(\ref{IR_condition_twisted}) is in general a necessary condition.
This is because our approach only depends on the information of the underlying theories, namely, massless fermions coupled to $\mathbb{Z}_n$/$\mathrm{spin}^{\mathbb{Z}_{2m}}$ gauge fields, while the Ib\'{a}\~{n}ez-Ross conditions rely on the anomaly constraints of embedding continuous gauge theories at UV and also on the symmetry breaking procedures from UV to IR.
\footnote{
From the aspect of the 't Hooft anomaly matching, one in principle needs to consider any possible embedding theories (not just U(1) gauge theories) at UV and any possible symmetry breaking procedures to obtain the complete anomaly constraints.
}
Therefore, our result gives a more fundamental understanding of anomalies for these discrete symmetries themselves than the argument in \cite{Ibanez-Ross1991}.

Nevertheless, one can still see some connection between (\ref{anomaly-free_Z_n})/(\ref{anomaly-free_twisted}) and (\ref{IR_condition_untwisted})/(\ref{IR_condition_twisted}). For the case of $\mathbb{Z}_n$ symmetry, it is easy to check that the condition (\ref{IR_condition_untwisted}) is actually identical to (\ref{anomaly-free_Z_n}). 
On the other hand, for the case of $\mathrm{spin}^{\mathbb{Z}_{2m}}$ symmetry with even $m$, there exists some sets of coefficients that (\ref{IR_condition_twisted}) is not consistent with (\ref{anomaly-free_twisted}). 
For example, for a theory of four left-handed Weyl fermions with the same $\mathrm{spin}^{\mathbb{Z}_{4}}$ charges $\tilde{s} =1 \mod 4$, (\ref{IR_condition_twisted}) is satisfied (for $p=p'=2$), while (\ref{IR_condition_untwisted}) is not.
It would be very interesting (though not simple) to show whether (\ref{IR_condition_twisted}) with some particular sets of coefficients $\{p, k\}$ is equal to the condition (\ref{anomaly-free_twisted}).

\paragraph{Trivialization of anomalies by symmetry extensions}
\

On the other hand, we know that the Ib\'{a}\~{n}ez-Ross conditions are subject to the issue of symmetry extensions, which is also crucial when considering the 't Hoot anomalies of discrete symmetries. That is, (some parts of the) anomalies can in general change or even disappear when symmetries are extended. This can be treated in a formal way as we consider the following group extension
\cite{Witten2016, Wang:2017aa}:
\begin{align}
\label{group_extension}
1 \rightarrow \mathcal{K} \rightarrow \mathscr{H} \rightarrow \mathscr{G} \rightarrow 1,
\end{align}
where the symmetry groups $\mathscr{G}$ ($\mathscr{H}$) can be $\mathrm{Spin}\times G$ or  $\mathrm{Spin}^{G}$ ($\mathrm{Spin}\times H$ or  $\mathrm{Spin}^{H}$) and all $G$, $H$, and $\mathcal{K}$ are finite. Given a homomorphism $\mathscr{H} \rightarrow \mathscr{G}$, if a nontrivial 4d anomaly (5d cobordism class) is pulled back to a trivial class or the identity element of $\Omega^5_{\mathscr{H}}$, we say that the anomaly $\beta$ is trivialized by extending $\mathscr{G}$ to $\mathscr{H}$ (via the above group extension).


For example, let us take $\mathscr{G}=\mathrm{Spin}\times\mathbb{Z}_4$, $\mathscr{H}=\mathrm{Spin}\times\mathbb{Z}_8$, $\mathcal{K}=\mathbb{Z}_2$ and an anomaly $e^{-2\pi i\alpha_R}\in\Omega^5_{\mathrm{Spin}}(B\mathbb{Z}_4)$ with $R=\rho_1\oplus\rho_1\oplus\rho_2\in RU(\mathbb{Z}_4)$. With the expression (\ref{anomaly_untwisted}) one can check that, for some suitable homomorphism $\mathbb{Z}_8\rightarrow\mathbb{Z}_4$, the pullback class $e^{-2\pi i\alpha_{R'}}$ with $R'=\rho_2\oplus\rho_2\oplus\rho_4\in RU(\mathbb{Z}_8)$   become trivial in $\Omega^5_{\mathrm{Spin}}(B\mathbb{Z}_8)$. This means that three left-handed Weyl fermions with symmetry transformations $\psi_{1,2}\rightarrow i\psi_{1,2}$ and $\psi_3\rightarrow -\psi_3$ cannot consistently couple to a $\mathbb{Z}_4$ gauge field because of a nonvanishing 't Hoot anomaly, but can couple to a $\mathbb{Z}_8$ gauge field  as the anomaly is trivialized.
If the $\mathbb{Z}_8$ gauge field is dynamical, such a (topological) gauge theory must support topological excitations (gapped degrees of freedom) transforming faithfully under $\mathbb{Z}_8$, so that the whole theory respects a $\mathrm{Spin}(4)\times\mathbb{Z}_8$ symmetry.

Similarly, one can also check that two Weyl fermions with a $\mathbb{Z}_4$ symmetry that $\psi_1\rightarrow i\psi_1$ and $\psi_2\rightarrow -\psi_2$ can never couple to a $\mathbb{Z}_n$ gauge theory via any symmetry extensions (that is, $n\in4\mathbb{N}$).  
Actually, under any group extension from $\mathbb{Z}_n$ to $\mathbb{Z}_{ln}$, $l>1$, the cubic term of the anomaly of $\mathrm{Spin}\times\mathbb{Z}_n$ always changes (and becomes trivial for some $l$), while the linear term can never been trivialized. 
The latter situation holds even for any symmetry extension from $\mathrm{Spin}\times \mathbb{Z}_n$ to $\mathrm{Spin}\times H$ with an arbitrary finite internal symmetry $H$. The linear anomaly of $\mathrm{Spin}\times G$, that is, the mixed $G$-gravitational anomaly, is represented by the 5d cobordism class $p_1\nu_G$, where $p_1$ is the first Pontryagin class and $\nu_G\in H^1(BG, U(1))=H^1(G, U(1))=\mathrm{Hom}(G, U(1))$. For any group extension 
$1 \rightarrow K \rightarrow H \overset{\pi}{\rightarrow} G \rightarrow 1$, a nontrivial class $\nu_G\in H^1(G, U(1))$, when pulled back to $\pi^*\nu_G\in H^1(H, U(1))$, can not be trivial. Identically, $p_1\nu_G\in\Omega^5_{\mathscr{G}}$ can not be trivialized in $\Omega^5_{\mathscr{H}}$ under the extension (\ref{group_extension}). 
\footnote{
The author thanks E. Witten for a useful discussion about this argument.
}
This is what exactly happens in the the Ib\'{a}\~{n}ez-Ross condition. 
Therefore, only the vanishing of the linear mixed anomaly should be required for the (effective) $\mathbb{Z}_n$ symmetry on massless fermions to be gauged, regardless of how the symmetry is extended when massive sector is included, as pointed out by Banks and Dine in \cite{Banks-Dine1991}.

We can also consider symmetry extensions from a twisted symmetry to an untwisted symmetry. For example, take $\mathscr{G}=\mathrm{Spin}^{\mathbb{Z}_4}$, $\mathscr{H}=\mathrm{Spin}\times\mathbb{Z}_4$, $\mathcal{K}=\mathbb{Z}_2$ and an anomaly $e^{-2\pi i\tilde{\alpha}_{\tilde{R}}}\in\Omega^5_{\mathrm{Spin}^{\mathbb{Z}_4}}$ with 
$\tilde{R}=\tilde{\rho}_1\oplus\tilde{\rho}_1\oplus\tilde{\rho}_1\oplus\tilde{\rho}_1\in RU^o(\mathbb{Z}_4)$. 
By (\ref{anomaly_untwisted}) and (\ref{anomaly_twisted}) we know that the pullback class $e^{-2\pi i\alpha_{\tilde{R}'}}$ with $\tilde{R}'=\rho_1\oplus\rho_1\oplus\rho_1\oplus\rho_1\in RU(\mathbb{Z}_4)$   become trivial in $\Omega^5_{\mathrm{Spin}}(B\mathbb{Z}_4)$. The physical meaning of such a trivialization is that four left-handed Weyl fermions with symmetry transformation $\psi_i\rightarrow i\psi_i$ cannot consistently couple to a $\mathrm{spin}^{\mathbb{Z}_4}$ gauge field, while there is no problem for them to couple to a $\mathbb{Z}_4$ gauge field.
In this situation, if the $\mathbb{Z}_4$ gauge field is dynamical, the corresponding (topological) $\mathbb{Z}_4$ gauge theory to which the massless fermions couple must be a spin TQFT (if not, the whole theory including both the massless and topological sectors can also be formulated on any non-spin manifold, which is a contradiction because of the presence of the $\mathrm{Spin}^{\mathbb{Z}_4}(4)$ anomaly).

%
%
%

\section{Massive fermions with anomalous global symmetries in 4d}
\label{Massive fermions with anomalous global symmetries in 4d}

We have studied the 't Hooft anomalies of $\mathrm{Spin}(4)\times G$ and $\mathrm{Spin}^{G}(4)$ symmetries in massless chiral fermion theories, focusing on the case $G$ is a cyclic group. 
Then one may ask: can massive fermions have a global symmetry enjoying an anomaly?
Note that a theory with a global symmetry possessing perturbative anomalies must be gapless, if the symmetry is not spontaneously broken.  However,  if a theory has a global symmetry with only non-perturbative anomalies, 
it is possible to have symmetry-respecting gapped ground states for such a system.
In this section, we present an approach, based on the idea in 
\cite{Seiberg-Witten2016, Witten2016, Wang:2017aa},
for constructing a gapped state of fermions with an anomalous global symmetry. It is convenient to think of this states as a boundary state of a nontrivial $4+1$-dimensional fermionic SPT phase. 
\footnote{
It is not necessary that such a gapped state must live on the boundary of some $(4+1)$d SPT phase, however, as an anomalous global symmetry can possibly be emergent at low energy in a purely $(3+1)$d system.
}


\subsection{Some information from $\Omega^{5}_{\mathscr{G}}$}
\label{Some information from the spin cobordism groups}

Before giving specific models of gapped states with anomalous global symmetries, let us see what we can know about the features of these states from the cobordism group $\Omega^{5}_{\mathscr{G}}$,
which classifies 4+1-dimensional SPT phases  with symmetry $\mathscr{G}$. 
In our case,  we have $\mathscr{G}=\mathrm{Spin}\times G$ or $\mathrm{Spin}^{G}$.
For a theory of fermions in any representation of $\mathscr{G}$ that corresponds to the identity element of  $\Omega^{5}_{\mathscr{G}}$, the bulk phase is topologically trivial 
and any associated boundary state is anomaly-free.
 In this case, a gapless boundary state can be turned into
a gapped, symmetry-preserving, and topologically trivial state, with the bulk phase (in the low-energy limit) unchanged.
Note that here we are allowed to add extra degrees of freedom (such as matter and gauge fields) and/or interactions which are only present on the boundary, that is, in a purely $(3+1)$d system, for constructing a gapped boundary state.  

On the other hand, a theory specified by a nontrivial (non-identity) element  $\Omega^{5}_{\mathscr{G}}$ has a nontrivial bulk phase and, while a boundary is present, a boundary state ---  not unique in general --- with an anomalous $\mathscr{G}$ symmetry. The standard boundary state consist of massless chiral fermions. We then want to know when such a boundary state can be gapped without symmetry breakdown on the boundary, which, again, can be studied by the idea of symmetry extension 
\cite{Witten2016, Wang:2017aa}.
An observation is that if there exists a nontrivial group extension (the same as (\ref{group_extension}))
\begin{align}
\label{group_extension_2}
1 \rightarrow \mathcal{K} \rightarrow \mathscr{H} \rightarrow \mathscr{G} \rightarrow 1,
\end{align}
where $\mathcal{K}$ is an emergent finite gauge group only coupled to boundary fermions, such that the pullback (associated with the homomorphism $ \mathscr{H} \rightarrow \mathscr{G}$) of the nontrivial element $\beta\in\Omega^{5}_{\mathscr{G}}$ becomes the identity element of $\Omega^{5}_{\mathscr{H}}$, then the  gapless boundary state can be driven to, by the gauge interaction associated to $\mathcal{K}$, a gapped state that respects a global symmetry $\mathscr{H}/\mathcal{K}\cong \mathscr{G}$. 
The gapless and the gapped boundary states have the same anomaly of $\mathscr{G}$, as they are both coupled to the same bulk SPT phase. Note, however, that such a gapped state is topologically nontrivial, since there is a $\mathcal{K}$ gauge symmetry present in the low energy boundary theory, and the (anomalous) global symmetry $\mathscr{G}$ is realized projectively on the boundary.

Let us look at an example, taking $\mathscr{G}=\mathrm{Spin}\times\mathbb{Z}_n$, $\mathcal{K}=\mathbb{Z}_l$, and  $\mathscr{H}=\mathrm{Spin}\times\mathbb{Z}_{ln}$ for left-handed spin 1/2 Weyl fermions, where $(l, n) \neq1$.
\footnote{
If $(l, n)=1$, $\mathbb{Z}_{ln}\cong\mathbb{Z}_l\times\mathbb{Z}_n$, so that (\ref{group_extension_2}) is a trivial extension that does not help us trivialize the cobordism class of $\Omega^{5}_{\mathscr{G}}$.
}
A nontrivial extension is specified by the following expressions for generators of these cyclic groups:
\begin{align}
\hat{S} &= \exp(2\pi i\hat{s}/n) \in \mathbb{Z}_n, \nonumber\\
\hat{K} &= \exp(2\pi i\hat{k}/m) \in \mathbb{Z}_l, \nonumber\\
\hat{H} &= \hat{S}\cdot \hat{K}^{1/n} = \exp(2\pi i (m\hat{s}+\hat{k})/mn) \in \mathbb{Z}_{ln},
\end{align}
where $\hat{s}$ and $\hat{k}$ are the (discrete) charge operators associated with $\mathbb{Z}_n$ and $\mathbb{Z}_l$ symmetries, respectively. Note that the generator of $\mathbb{Z}_{ln}$ satisfies $\hat{H}^n = \hat{S}^n\cdot\hat{K}=\exp(2\pi i (\hat{s}+\hat{k}/m)) \in \mathbb{Z}_m$, so we indeed have $\mathbb{Z}_{ln}/\mathbb{Z}_l\cong\mathbb{Z}_n$ (and thus $\mathscr{H}/\mathcal{K}\cong \mathscr{G}$). Now, we would like to know when a nontrivial element $\exp(-2\pi i\alpha_{R})\in\Omega^{5}_{\mathrm{Spin}}(B\mathbb{Z}_n)$, with a representation $R=\oplus_i e^{2\pi i s_i/n}$,  can be trivialized in $\Omega^{5}_{\mathrm{Spin}}(B\mathbb{Z}_{ln})$. The pullback operation gives
\begin{align}
\label{pullback_element}
\alpha_R &= \left(\frac{1}{6n}\left(n^2+3n+2\right)\sum_i  s_i^3 \mod\mathbb{Z},\quad \frac{2}{n}\sum_i s_i \mod\mathbb{Z} \right)
\nonumber\\
&\rightarrow
\left(\frac{1}{6ln}\left(l^2n^2+3ln+2\right)\sum_i  (ls_i+k_i)^3 \mod\mathbb{Z},\quad \frac{2}{ln}\sum_i (ls_i+k_i) \mod\mathbb{Z} \right).
\end{align}

As $\mathcal{K}$ is a gauge group that only appears on the boundary of a $(4+1)$d fermionic SPT phase,
there should be constraints on the $\mathbb{Z}_l$ charges $\{k_{i}\}$, which are,
by looking at (\ref{pullback_element}), 
\begin{align}
\sum_i  \left[(ls_i+k_i)^3 - (ls_i)^3\right] = 0 \mod a_{ln},
\quad\sum_i k_i= 0 \mod ln/2,
\end{align}
where $a_{ln}$ is defined in (\ref{def_an_bn}).
Under these constraints, a solution for $l$ and $\{k_{i}\}$ that makes the pullback element in (\ref{pullback_element}) trivial exists only when 
\begin{align}
\sum_i s_i= 0 \mod n/2, 
\end{align}
and one can take, for example,  $l^3=0 \mod a_{ln}$ as a solution. 
It is noted that, if $\sum_i s_i^3 \neq 0 \mod a_{n}$, it is impossible to trivialize $\alpha_R$ when $m$ and $n$ are coprime. This is why we need to consider a nontrivial group extension for trivializing a boundary anomaly. 

As discussed in the last section, only a $\mathbb{Z}_n$ symmetry with vanishing linear (mixed) anomaly can be gauged, up to symmetry extensions. The same argument applies here: only massless fermions in a representation of $\mathbb{Z}_n$ with vanishing linear anomaly can be gapped, through a symmetry extension $\mathbb{Z}_n$ to $H$ (not necessarily a cyclic group), in a symmetry-preserving manner.


In the next section,  we give a physical model to realize gapped states with an anomalous global $\mathbb{Z}_n$ symmetry. The idea presented here will be more concrete.

\subsection{A model via weak coupling}
\label{A model via weak coupling}

Now we present a model of gapped states of fermions with an anomalous global (untwisted) $\mathbb{Z}_n$ symmetry, in the framework of weak coupling. Our approach is similar to 
that in \cite{Ibanez-Ross1991} for deriving the Ib\'{a}\~{n}ez-Ross condition for a $\mathbb{Z}_n$ gauge symmetry (reviewed in Sec.~\ref{Review of the Ib\'{a}\~{n}ez-Ross conditions for discrete symmetry anomalies}), and is a $3+1$-dimensional analog of that in \cite{Seiberg-Witten2016, Witten2016} for constructing gapped boundary states in $2+1$ dimensions. 

According to the analysis in Sec.~\ref{Some information from the spin cobordism groups}, one way to construct a symmetric gapped state is to lift the global symmetry $\mathbb{Z}_n$ to  $\mathscr{H}$ by a gauge group $\mathcal{K}$. To do this, one can begin with a trivial group extension of $\mathbb{Z}_n$ by a
U(1) gauge group
\begin{align}
1 \rightarrow U(1) \rightarrow U(1)\times \mathbb{Z}_n \rightarrow \mathbb{Z}_n \rightarrow 1,
\nonumber
\end{align}
and then breaks $U(1)\times\mathbb{Z}_n$ down to $\mathbb{Z}_{ln}$ spontaneously at low energy.
We first have to determine in what representation of U(1), while we are given a set of chiral fermions (on the boundary) in a representation of $R$ of  $\mathbb{Z}_n$, there is no extra gauge anomaly and only the $\mathbb{Z}_n$ anomaly (represented by $\alpha_R$ in (\ref{anomaly_untwisted})) is present. In general, this is not easy to compute; besides the gauge and mixed gauge-gravitational anomalies for U(1) itself, we also have to take the mixed anomalies between U(1) and $\mathbb{Z}_n$ into account. 
(That is, we need to know the full 't Hooft anomaly of the $\mathrm{Spin}(4)\times\mathbb{Z}_n\times\mathrm{U}(1)$ group.)
Instead of looking at the most general case, we consider a representation
\begin{align}
\mathbb{1}_{\mathrm{U}(1)}\otimes R \oplus \mathcal{R}\otimes \mathbb{1}_{\mathbb{Z}_n},
\end{align}
where $\mathcal{R}$ is a representation of U(1) and $\mathbb{1}_{\mathrm{U}(1)}$ (with the dimension equal to $R$) and $\mathbb{1}_{\mathbb{Z}_n}$ (with the dimension equal to $\mathcal{R}$)  are respectively trivial representations of U(1) and $\mathbb{Z}_n$. To make sure that such a representation has the same $\mathbb{Z}_n$ anomaly as $R$, we only need to check there is no (perturbative) gauge and mixed gauge-gravitational anomalies for U(1). 

Specifically, let $\{\psi_i\}$ and $\{\chi_{j}\}$ be two sets of left-handed Weyl fermions transforming in the representations $\mathbb{1}_{\mathrm{U}(1)}\otimes R$ and $\mathcal{R}\otimes \mathbb{1}_{\mathbb{Z}_n}$ of the group $\mathrm{U}(1)\times\mathbb{Z}_n$, respectively. 
Let $R=\oplus_i \rho_{s_i}$, where $\rho_{s_i}$ is a one-dimensional representation with a $\mathbb{Z}_n$ charge $s_i$, 
and $\{k_j\in\mathbb{Z}\}$ be the U(1) charges associated to $\mathcal{R}$, which satisfy the anomalies constraints $\sum_j k_j^3 = 0$ and $\sum_j k_j= 0$.
The Lagrangain of the fermions coupled to an emergent U(1) gauge field $a$ that propagates only along the boundary is given by 
\begin{align}
L_0=
\sum_ i\overline{\psi}_{i}i\slashed{D}_0P_+\psi_{i}+
\sum_j \overline{\chi}_{j}(i\slashed{D}_0+k_j\slashed{a})P_+\chi_{j},
\end{align}
where $P_+=(1+\gamma_5)/2$ and $\slashed{D}_0$ is the Dirac operator for a fermion coupled to gravity only.
In the absence of the U(1) gauge symmetry, the fermions $\{\chi_j\}$ can be fully gapped --- as each of them can have a Majorana mass $m_j\overline{\chi}_{j}\chi_{j}^c$ --- and thus the low energy theory is just the standard boundary state, described by the massless chiral fermions $\{\psi_i\}$, of a $(4+1)$d fermionic $\mathbb{Z}_n$ SPT phase represented by $\alpha_R$ in (\ref{anomaly_untwisted}).

\begin{figure}[tbp]
\centering
\scalebox{0.7}{\includegraphics{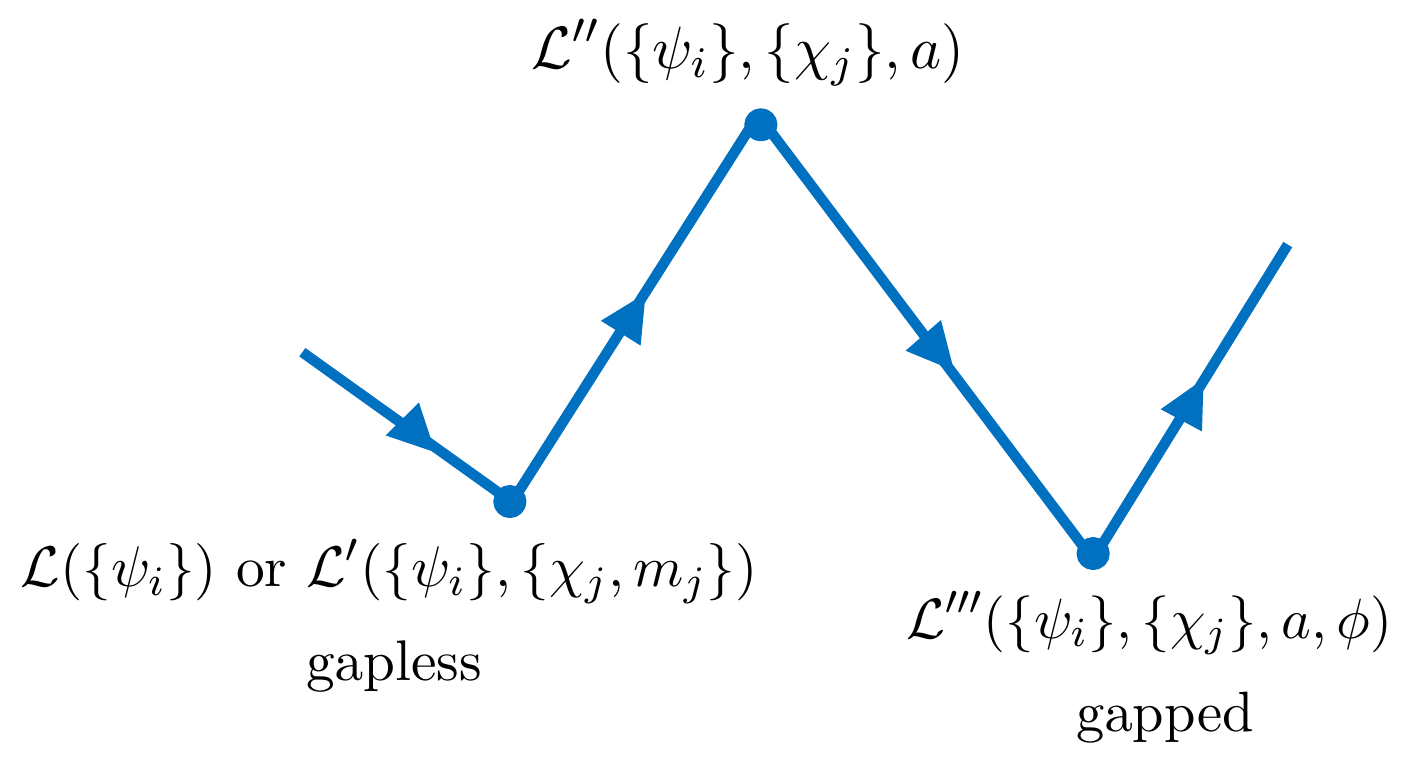}}
\caption{
\label{RG_BoundaryStates}
Constructing different boundary theories of the same SPT phase by changing the degrees of freedom on the boundary.
}

\end{figure}

In the presence of the U(1) gauge symmetry, the usual mass terms are forbidden, and the part of $L_0$ that includes the $\chi_j$ fermions describes a chiral gauge theory. Nevertheless, one can introduce Higgs fields (charge scalar fields) and Yukawa couplings among $\psi_i$ and $\chi_j$ fermions, so that all the fermions receive masses from the expectation values of the Higgs fields. Here we consider a one-Higgs model with the following (nonlinear) Yukawa couplings
\begin{align}
L_{\mathrm{Yuk}}
= \sum_{\mathrm{pairs}\{i, i'\},\ \mathrm{pairs}\{j', j''\},\ l} \left\{ 
\lambda_{i,i'}\phi^{\alpha_{i,i'}} \overline{\psi}_{i}\chi_{i'}^c
+ g_{j', j''}\phi^{\beta_{j', j''}} \overline{\chi}_{j'}\chi_{j''}^c
+ h_{l}\phi^{\gamma_{l}}\overline{\chi}_{l}\chi_{l}^c
\right\}+\mathrm{h.c.},
\end{align}
for some coupling constants $\lambda_{i, i'}$, $g_{j', j''}$, and $h_l$.
Here $\alpha_{i, i'}$, $\beta_{j', j''}$, and $\gamma_l$ are nonnegative integers
and $\chi_j^c := i\gamma^0 \mathcal{C}\chi_j^*$, with $\mathcal{C}$ being the charge conjugation matrix, are the charge conjugate fields of $\chi_j$ and right-handed.
We have also divided the set of fermions $\{\chi_j\}$ into four distinct sets $\{\chi_{i'}\}$, 
$\{\chi_{j'}\}$, $\{\chi_{j''}\}$, and $\{\chi_{l}\}$ (here we assume the number of $\chi_j$ is not smaller than the number of $\psi_i$), such that each $\chi_{i'}$ is paired up with each $\psi_i$ with a Dirac-type mass,  each $\chi_{j'}$ is paired up with each $\chi_{j''}$ with a Dirac-type mass, and each $\chi_{l}$ itself is with a Majorana-type mass. 
(A flows connecting theories with different degrees of freedom but the same anomaly of $\mathbb{Z}_n$ is schematically shown in Figure. \ref{RG_BoundaryStates}.)
Denote the $\mathbb{Z}_n$ and the U(1) charges of $\phi$ be $\bar{s}$ and $\bar{k}$, respectively. Now, we require $L_{\mathrm{Yuk}}$ to be invariant under both the $\mathbb{Z}_n$ global symmetry and the U(1) gauge symmetry. This constrains the values of the $\mathbb{Z}_n$ and the U(1) charges of all particles:
\begin{align}
\text{Invariant under $\mathbb{Z}_n$}&: \quad
\alpha_{i, i'}\bar{s}-s_i \in n\mathbb{Z}, \quad\beta_{j', j''}\bar{s} \in n\mathbb{Z}, \quad\gamma_{l}\bar{s} \in n\mathbb{Z}.
\nonumber\\
\text{Invariant under U(1)}&: \quad
\alpha_{i, i'}\bar{k}=k_{i'}, \quad\beta_{j', j''}\bar{k}=k_{j'}+k_{j''}, \quad\gamma_{l}\bar{k}=2k_{l}.
\end{align}
Substituting the above conditions to the anomalies constraints on $k_j$, that is, $\sum_j k_j^3 = 0$ and $\sum_j k_j= 0$, we have
\begin{align}
\label{gapped_states_condition_Z_n}
\sum_i (\bar{k}s_i)^3 &= p\cdot\bar{k}n+ r\cdot\frac{(\bar{k}n)^3}{8}, \quad p, r\in\mathbb{Z};\ p \in 3\mathbb{Z}\ \text{if}\ n\in 3\mathbb{Z},
 \nonumber\\
\sum_i s_i &= p'\cdot n+ r'\cdot \frac{n}{2}, \quad p', r'\in\mathbb{Z},
\end{align}
which can also be written, in terms of $a_{n}$ defined in (\ref{def_an_bn}), as
\begin{align}
\label{gapped_states_condition_Z_n_2}
\sum_i (\bar{k}s_i)^3 = k a_{\bar{k}n},
\quad
\sum_i s_i= \ell n/2,
\quad
k, \ell \in\mathbb{Z}.
\end{align}

When the field $\phi$ has a nonzero expectation value $\langle\phi\rangle$, the theory becomes gapped, and the U(1) gauge group is broken down to a finite subgroup $\mathbb{Z}_{\bar{k}}$, as $\phi$ carries a U(1) charge $\bar{k}$. On the other hand, the original (microscopic) $\mathbb{Z}_n$ global symmetry is also broken, since $\langle\phi\rangle$ is not invariant under a generator $\hat{S}\in\mathbb{Z}_n$ (as $\bar{s}$ is in general not equal to 0 modulo $n$). 
For convenience, let us assume $\bar{s}= - 1$.
Nevertheless, such a gapped phase respects another symmetry which is the combination of $\hat{S}$ with a gauge symmetry $\hat{K}^{1/n}$ (which is also broken by $\langle\phi\rangle$), where $\hat{K}$ is a generator of the low energy $\mathbb{Z}_{\bar{k}}$ gauge group, and we denote it as
\begin{align}
\hat{H} &= \hat{S}\cdot \hat{K}^{1/n} = \exp(2\pi i (\bar{k}\hat{s}+\hat{k})/\bar{k}n) \in\mathbb{Z}_{\bar{k}n},
\end{align}
where $\hat{s}$ and $\hat{k}$ are the charge operators associated with $\mathbb{Z}_n$ and U(1), respectively. The definition of $\hat{H}$ is not unique; any operator having the form $\hat{H}\cdot\hat{K}'$ for any $\hat{K}'\in\mathbb{Z}_{\bar{k}}$ is also a symmetry of the low energy phase.
The point is that $\hat{H}$ (or any $\hat{H}\cdot\hat{K}'$) has the following property
\begin{align}
\hat{H}^n = \hat{S}^n\cdot \hat{K} \in \mathbb{Z}_{\bar{k}},
\end{align}
so that any physical state invariant under the low energy $\mathbb{Z}_{\bar{k}}$ gauge symmetry satisfies $\hat{H}^n=1$. 
On the other hand, individual gauge non-invariant quasiparticles transform under a $\mathbb{Z}_{\bar{k}n}$ symmetry, so the low energy gapped phase respects a global $\mathbb{Z}_n$ symmetry that is realized projectively in the presence of the $\mathbb{Z}_{\bar{k}}$ gauge symmetry

Therefore, if we are given a set of Weyl fermions $\{\psi_i\}$ with $\mathbb{Z}_n$ charges $\{s_i\}$ obeying the condition (\ref{gapped_states_condition_Z_n}) or (\ref{gapped_states_condition_Z_n_2}) for some $\bar{k}\in\mathbb{Z}$, it is possible, using the model presented here, to construct a gapped state of fermions that also preserves a global $\mathbb{Z}_n$ symmetry. If there exists a solution for (\ref{gapped_states_condition_Z_n}) or (\ref{gapped_states_condition_Z_n_2}) with $\bar{k}=1$, such a symmetric gapped state is topologically trivial. On the other hand, if one can only find a solution for (\ref{gapped_states_condition_Z_n}) or (\ref{gapped_states_condition_Z_n_2}) with some integer $\bar{k}\neq 1$, the corresponding gapped state would be topologically nontrivial, as there is a $\mathbb{Z}_{\bar{k}}$ gauge symmetry emergent at low energy (while the full symmetry of the system is lifted from $\mathbb{Z}_n$ to $\mathbb{Z}_{\bar{k}n}$ by this $\mathbb{Z}_{\bar{k}}$ gauge symmetry). 
This gapped state has an anomalous $\mathbb{Z}_n$ global symmetry. 

The physical model of gapped (boundary) states via weak coupling we considered agrees with the analysis by purely geometrical considerations --- knowledge from the cobordism groups $\Omega^{5}_{\mathrm{Spin}}(B\mathbb{Z}_n)$ --- in the last section, as expected. In this model, however, we are not sure if there always exists a set of Weyl fermions $\{\chi_j\}$ in an anomaly-free U(1) representation such that (\ref{gapped_states_condition_Z_n}) or (\ref{gapped_states_condition_Z_n_2}) has a solution for each given $\{\psi_i\}$ with $\mathbb{Z}_n$ charges $\{s_i\}$ that satisfy this condition. The same situation also happens to the Ib\'{a}\~{n}ez-Ross condition for an anomaly-free $\mathbb{Z}_n$ symmetry.
\newline

\paragraph{Examples}
\

Let us look at some examples for the construction of gapped boundary states. Consider $\nu$ left-handed Weyl fermions $\{\psi_i\}$ in a representation $R=\oplus_{i=1}^{\nu} e^{2\pi i s_i/4}$ of a $\mathbb{Z}_4$ global  symmetry. Since we have $\alpha_{\rho_{2}}=0$ and $\alpha_{\rho_{3}}= -\alpha_{\rho_{1}}$ (without symmetry breaking, a Weyl fermion with $\mathbb{Z}_4$ charge 2 can be gapped by a Majorana mass, while a pair of Weyl fermions with $\mathbb{Z}_4$ charges 1 and 3 can be gapped by a Dirac mass), we can focus on the case where all $s_i=1$:

(1) For $\nu=0 \mod 4$,  $\alpha_R=(\nu/4 \mod\mathbb{Z},\ 0 \mod\mathbb{Z}) = 0$ and thus the theory is anomaly-free.
Let us take $\nu=4$ for discussion. To construct a gapped, symmetry-preserving, and topological trivial state by the model above, we can take, for example, the U(1) charges of $\{\chi_j\}$ as 
$\{k_j\}=\{3, -5, -5, -5, -1, 5, 2, 6\}$, such that $\chi_{i'}$ and $\psi_i$ for $i'=i=1,..., 4$, $\chi_5$ and $\chi_6$, and $\chi_7$ and $\chi_8$ are all paired with Dirac-type masses
(there are no fermions with Majorana-type masses in this case), and also take the $\mathbb{Z}_4$ and the U(1) charges $\bar{s}$ and $\bar{k}$ of $\phi$ to be $-1$ and $1$, respectively.
The corresponding low energy phase in the presence of a nonzero $\langle\phi\rangle$ is invariant under a $\mathbb{Z}_4$ global symmetry 
$\hat{H}=  \exp(2\pi i (\hat{s}+\hat{k})/4)$ that comes from a breakdown of the (high energy) $U(1)\times\mathbb{Z}_4$ symmetry.

(2) For $\nu=2 \mod 4$,  $\alpha_R=(1/2 \mod\mathbb{Z},\ 0 \mod\mathbb{Z}) \neq 0$, while the linear anomaly $\sum_is_i /2\mod \mathbb{Z}$ is trivial.  
In this case, we can have a gapped state with an anomalous $\mathbb{Z}_4$ symmetry.
Take $\nu=2$ for discussion. Let $\{k_j\}=\{-2, -10, 7, 9, -4\}$, $\bar{s}=-1$, and $\bar{k}=2$, so that $\chi_1$ and $\psi_1$, $\chi_2$ and $\psi_2$, and $\chi_3$ and $\chi_4$ are all paired with Dirac-type masses, while the fermion $\chi_5$ is with a Majorana-type mass. The low energy phase has a $\mathbb{Z}_2$ gauge symmetry and we can define a $\mathbb{Z}_4$ global symmetry via a symmetry $\hat{H}=  \exp(2\pi i (2\hat{s}+\hat{k})/8)\in\mathbb{Z}_8$, as  any state of a compact sample satisfies $\hat{H}^4=\exp(\pi i \hat{k}) =1$. Note that individual quasiparticles such as $\chi_3$ and $\chi_4$ have a ``symmetry-fractionalization'' relation $\hat{H}^4= -1$.

(3) For odd $\nu$,  $\alpha_R \neq 0$, and the linear anomaly also does not vanish. From the previous discussion, we have known it is impossible to trivialize $\alpha_R$ by extending $\mathbb{Z}_4$ to any symmetry group $H$ (not just the cyclic group), and thus we can not have a gapped state from any model associated with this kind of symmetry extensions.

%
%
%
%
%
%
%

\section{Summary}
\label{Summary}

In this work, we compute the 't Hooft anomalies of the $\mathrm{Spin}(4)\times\mathbb{Z}_n$ and the $\mathrm{Spin}^{\mathrm{Z}_{2m}}(4)=(\mathrm{Spin}(4)\times\mathbb{Z}_{2m})/\mathbb{Z}_{2}$ symmetry group in a $(3+1)$-dimensional chiral fermion theory. These anomalies are identified as elements of the five-dimensional cobordism groups associated with these symmetries, and we give expressions of them in terms of the $\mathbb{Z}_n$ and the $\mathrm{spin}^{\mathrm{Z}_{2m}}$ representations of fermions, as shown in (\ref{anomaly_untwisted}) and (\ref{anomaly_twisted}), respectively. This gives us the anomaly-free conditions (\ref{anomaly-free_Z_n}) and (\ref{anomaly-free_twisted}). In particular, (\ref{anomaly-free_Z_n}) is identical to the Ib\'{a}\~{n}ez-Ross condition condition (\ref{IR_condition_untwisted}), which is deduced by anomaly matching between the low energy $\mathbb{Z}_n$ symmetry and an embedding U(1) symmetry at higher energy scale.

For any consistent chiral gauge theory --- Weyl fermions coupled a topological gauge theory --- with a definite  full symmetry group $\mathrm{Spin}(4)\times\mathbb{Z}_n$ or $\mathrm{Spin}^{\mathrm{Z}_{2m}}(4)$, the discrete charges of the massless Weyl fermions must \emph{strictly} satisfy the condition (\ref{anomaly-free_Z_n}) or (\ref{anomaly-free_twisted}). However, if only the (effective) symmetry on the massless fermions is known and no information about (the symmetry on) the massive degrees of freedom or the topological gauge theory is specified, these anomaly constrains for the massless fermions should only be respected \emph{up to} symmetry extensions.

We also apply the idea of anomaly trivialization by symmetry extensions to study symmetric gapped states of fermions with an anomalous $\mathbb{Z}_n$ symmetry. A simple way to construct these gapped states is by a weakly coupled model. Finding a TQFT description of these states (in the low energy limit) is left for future work. It would also be interesting if (some of) these nontrivial gapped states can be realized, with an emergent anomalous global symmetry, in the low energy phase of a physical system.

%
%
%

\acknowledgments
The author thanks Miguel Montero, Hitoshi Murayama, Shinsei Ryu, Yuji Tachikawa, Edwar Witten, and Kazuya Yonekura for helpful discussions. The author thanks Yuji Tachikawa for useful comments on the draft, and also thanks Miguel Montero for sharing their (in collaboration with I\~{n}aki Garc\'{i}a-Etxebarria) manuscript before publication. 
This work was supported by World Premier International Research Center Initiative (WPI), MEXT, Japan.

\appendix


\section{Details of some derivations}
\label{Details of some derivations}


\subsection{Derivation of Eq.\ (\ref{Sn_2})}
\label{isomorphisms}

Here we follow the idea in 
\cite{BarreraYanes-Gilkey1999} for the case $n=2^v$
and generalize their result to any prime power $n=p^v$.
Eq.\ (\ref{Sn_2}) can be proved by relating the $\eta$ invariant on an element of $S_n$ to the $\eta$ invariant on a seven-dimensional lens space $L(n; 1, 1, 1, a)$. 
Here 
\begin{align}
L(n; a_1, a_2, a_3, a_4) := S^7/ \tau(a_1, a_2, a_3, a_4),
\end{align}
where $\tau(a_1, a_2, a_3, a_4) := \rho_{a_1}\oplus\rho_{a_2}\oplus\rho_{a_3}\oplus\rho_{a_4}$ is a representation of $\mathbb{Z}_n$ in $\mathrm{U}(4)$ and its action (by multiplication by $\lambda^{a_i}$ on the $i$-th summand) on the associated unit sphere bundle is fixed-point free. 
By construction, the lens spaces inherit natural spin structures and $\mathbb{Z}_n$ structures 
(similar to the case of lens space bundles $X(n; a_1, a_2)$). 
The $\eta$ invariant on a generic ($4j-1$)-dimensional lens space $L(n; a_1, ..., a_{2j})$ --- with the natural spin$\times\mathbb{Z}_n$ structure --- in a representation $R\in RU(\mathbb{Z}_n)$ can be computed by the following formula 
\cite{Botvinnik-Gilkey-Stolz1997, BarreraYanes-Gilkey1999}
\begin{align}
\label{combinatorial_eta_lens_space_spinxZn}
\eta(L(n; a_1, a_2, ..., a_{2j}), R) =  \frac{1}{n} \sum_{\lambda\in\mathbb{Z}_n, \lambda\neq 1} \mathrm{Tr}(R(\lambda))
\frac{\lambda^{\frac{1}{2}(a_1+a_2+\ \cdots\ +a_{2j})}}{(1-\lambda^{a_1})(1-\lambda^{a_2})\cdots (1-\lambda^{a_{2j}})}.
\end{align}

Let
\begin{align}
\label{Tn_1}
T_n :=
\left\{
\begin{array}{ll}
\mathrm{span}_{\mathbb{Z}}\{X(n; 1, 1)\},
\quad&\text{if}\ n=2, 3,
 \\
\mathrm{span}_{\mathbb{Z}}\{X(n; 1, 1), X(n; 1, 3)\},
\quad &\text{if}\ n=2^v> 2,
\\ 
\mathrm{span}_{\mathbb{Z}}\{X(n; 1, 1), X(n; 1, 5)\},
\quad &\text{if}\ n=3^v>3,
\\ 
\mathrm{span}_{\mathbb{Z}}\{X(n; 1, 1), X(n; 1, 3)\},
\quad &\text{if}\ n=p^v,\  p>3.
 \end{array}
\right.
\end{align}
Note that elements of $T_n$ are manifolds (together with all the relevant structures), while elements of $S_n$ defined in (\ref{Sn_1}) are equivalent classes of manifolds in $\Gamma^{\mathrm{Spin}}_5(B\mathbb{Z}_n)$.  Let $\gamma := \rho_1-\rho_{-1}$ and $\xi := \rho_{-1}(\rho_0-\rho_1)^2$.
We define an additive map $\sigma_n:  T_n \rightarrow  RU_0(\mathbb{Z}_n)$ for each $n=p^v$ via the following relations on generators of $T_n$:
\begin{align}
\label{def_sigma_n}
\left\{
\begin{array}{ll}
\sigma_n(X(n; 1, 1))=\gamma,
\quad&\text{if}\ n=2, 3,
 \\
\sigma_n(X(n; 1, 3))=\gamma,\quad
\sigma_n(X(n; 1, 1)-3X(n; 1, 3))=\gamma\xi,
\quad &\text{if}\ n=2^v>2,
\\ 
\sigma_n(X(n; 1, 5))=\gamma,\quad
\sigma_n(X(n; 1, 1)-5X(n; 1, 5))=5\gamma\xi+\gamma\xi^2,
\quad &\text{if}\ n=3^v>3,
\\ 
\sigma_n(X(n; 1, 3))=\gamma,\quad
\sigma_n(X(n; 1, 1)-3X(n; 1, 3))=\gamma\xi,
\quad &\text{if}\ n=p^v,\  p>3.
 \end{array}
\right.
\end{align}

Then, for any $X\in T(n)$ (given $n=p^v$) and for any $R\in RU(\mathbb{Z}_n)$, we have
\begin{align}
\label{connection_eta_5d_7d}
\eta(X, R)= \eta(L(n; 1, k_n, 1, -1), \sigma_n(X)R),
\end{align}
where
$k_n=1$, if $n=2,3$; $k_n=3$, if $n=p^v>2,\ p\neq 3$; $k_n=5$, if $n=3^v>3$.
Eq.\ (\ref{connection_eta_5d_7d}) can be checked directly using the formulas of the $\eta$ invariants on the lens space bundles and on the lens spaces, that is, Eqs.\ (\ref{combinatorial_eta}) and (\ref{combinatorial_eta_L}). Here we skip the computation details.

For each $n=p^v$, if $[X]_{\eta}=0$ in $\Gamma^{\mathrm{Spin}}_5(B\mathbb{Z}_n)$, we will have 
$\eta(X, R) \in\mathbb{Z}$ for all $R\in RU(\mathbb{Z}_n)$, by the definition of $\Gamma^{\mathrm{Spin}}_5(B\mathbb{Z}_n)$. From (\ref{connection_eta_5d_7d}), we then have 
$\eta(L(n; 1, k_n, 1, -1), \sigma_n(X)R) \in\mathbb{Z}$ for all $R\in RU(\mathbb{Z}_n)$.
This implies $\sigma_n(X)\in RU_0(\mathbb{Z}_n)^4$ for 
$n=2^v$ \cite{BarreraYanes-Gilkey1999} and for 
$n=p^v$ with $p$ being an odd prime \cite{Gilkey}. 
Therefore, $\sigma_n$ induces a well-defined map from $S(n)$ to $RU_0(\mathbb{Z}_n)/RU_0(\mathbb{Z}_n)^4$.

Now we show that the (induced) map $\sigma_n$ is an isomorphsim from $S_n$ to $I_n/\{I_n\cap RU_0(\mathbb{Z}_n)^4\}$ for each $n=p^v$. We argue this as follows.
If $\sigma_n([X]_{\eta})=0$ (in $RU_0(\mathbb{Z}_n)/RU_0(\mathbb{Z}_n)^4$), from (\ref{connection_eta_5d_7d}) we have $\eta(X, R)\in\mathbb{Z}$ for all $R\in RU(\mathbb{Z}_n)$. Again, by the definition of $\Gamma^{\mathrm{Spin}}_5(B\mathbb{Z}_n)$,  $[X]_{\eta}=0$ in $\Gamma^{\mathrm{Spin}}_5(B\mathbb{Z}_n)$ and, of course, in $S_n$. So $\sigma_n$ is injective.
On the other hand, the set $I_n$ defined in (\ref{In}) is generated by $\rho_s - \rho_{-s}$ for $s=0,..., n-1$ or equivalently by $\gamma\xi^j$ for $j\ge 0$. 
It is obvious that $\sigma_n$ for $n=2,3,4$ is surjective, since $I_2=0$ and $I_3=I_4= \gamma\cdot\mathbb{Z}$. For $n=p^v > 4$, $I_n/\{I(_n\cap RU_0(\mathbb{Z}_n)^4\}$ is generated by $\gamma$ and $\gamma\xi$ only (as $\gamma\xi^j$ for $j\ge2$ is 0 modulo $RU_0(\mathbb{Z}_n)^4$), so $\sigma_n$ is surjective from $S_n$ to $I_n/\{I_n\cap RU_0(\mathbb{Z}_n)^4\}$.
This completes the proof of (\ref{Sn_2}).

\subsection{Derivation of Eq.\ (\ref{eta_5d_7d_all_n})}
\label{5d eta invariants from 7d}

The equation (\ref{connection_eta_5d_7d}) can also be expressed as 
\begin{align}
\label{connection_eta_5d_7d_2}
\eta(X, R)=  \eta(L(n; 1, 1, 1, 1), h_n\cdot\sigma_n(X)R),
\end{align}
where  $h_n$ for each $n=p^v$ is an automorphism on  $I_n$ that is not in $RU_0(\mathbb{Z}_n)$. From the argument in \ref{isomorphisms}, we know $\tau_n := h_n\cdot\sigma_n$ also induces an isomorphism between $S_n$ and $I_n/\{I_n\cap RU_0(\mathbb{Z}_n)^4\}$, that is, 
\begin{align}
\eta([X], R)=\eta(L(n; 1, 1, 1, 1), \tau_n([X])R)  \mod\mathbb{Z},
\quad n=p^v.
\end{align}

The above relation can be extended to any positive integer; for any given $n\in\mathbb{N}$, it is always possible to find a set (as $S_n$) whose elements are lens space bundles and to construct a map (as $\sigma_n$ or $\tau_n$) from such a set to $I_n$ that is an isomorphism. Such a set, as agued in the main context,  is identical to $\Gamma^{\mathrm{Spin}}_5(B\mathbb{Z}_n)$ and $\Omega^{\mathrm{Spin}}_5(B\mathbb{Z}_n)$.

\bibliographystyle{JHEP}
\bibliography{references}

\end{document}